\documentclass{sig-alternate}
\usepackage[usenames,dvipsnames]{color}
\usepackage{url}
\usepackage{graphicx}
\usepackage{subfigure}
\usepackage{comment}
\usepackage{times}
\usepackage{algorithmic}
\usepackage{algorithm, array}

\begin{document}
\conferenceinfo{KDD'12,} {August 12--16, 2012, Beijing, China.}
\CopyrightYear{2012}
\crdata{978-1-4503-1462-6 /12/08}
\clubpenalty=10000
\widowpenalty = 10000

\definecolor{BLUE}{rgb}{0,0,1}

\newcommand{\hide}[1]{}
\newcommand{\note}[1]{\textcolor{black}{Note: #1}}
\newcommand{\rev}[1]{{\bf\MakeUppercase{[#1]}}}
\newcommand{\cz}[1]{{\bf\MakeUppercase{[#1]}}}
\newcommand{\sanjay}[1]{\textcolor{blue}{SK: #1}}
\newcommand{\dan}[1]{\textcolor{magenta}{DW: #1}}
\newcommand{\xhdr}[1]{\vspace{2mm}\noindent{{\bf #1.}}}
\newcommand{\eg}{\emph{e.g.}}
\newcommand{\ie}{\emph{i.e.}}
\def\imagetop#1{\vtop{\null\hbox{#1}}}
\newcolumntype{M}{>{$\vcenter\bgroup\hbox\bgroup}c<{\egroup\egroup$}}
 
\newcommand{\denselist}{ \itemsep -5pt\topsep-10pt\partopsep-10pt }
\newcommand{\denselistRefs}{ \itemsep -2pt\topsep-5pt\partopsep-7pt }

\title{Information Diffusion and External Influence in Networks
}

\numberofauthors{3}
\author{
\alignauthor Seth Myers\\
\affaddr{Stanford University}\\
\email{samyers@stanford.edu}\\
\alignauthor Chenguang Zhu  \\
\affaddr{Stanford University}\\
\email{cgzhu@stanford.edu}\\
\alignauthor Jure Leskovec\\
\affaddr{Stanford University}\\
\email{jure@cs.stanford.edu}\\
}

\maketitle

\begin{abstract}
Social networks play a fundamental role in the diffusion of information. However, there are two different ways of how information reaches a person in a network. Information reaches us through connections in our social networks, as well as through the influence of external out-of-network sources, like the mainstream media. While most present models of information adoption in networks assume information only passes from a node to node via the edges of the underlying network, the recent availability of massive online social media data allows us to study this process in more detail.

We present a model in which information can reach a node via the links of the social network or through the influence of external sources. We then develop an efficient model parameter fitting technique and apply the model to the emergence of URL mentions in the Twitter network.
%
Using a {\em complete} one month trace of Twitter we study how information reaches the nodes of the network.  We quantify the external influences over time and describe how these influences affect the information adoption. We discover that the information tends to ``jump'' across the network, which can only be explained as an effect of an unobservable external influence on the network. We find that only about 71\% of the information volume in Twitter can be attributed to network diffusion, and the remaining 29\% is due to external events and factors outside the network.



\end{abstract}

\noindent
{\bf Categories and Subject Descriptors:} H.2.8 {\bf [Database
Management]}: {Database Applications -- {\em Data mining}}

\noindent
{\bf General Terms:} Algorithms, theory, experimentation.

\vspace{-.1cm}

\section{Introduction}

\label{sec:intro}



\begin{figure}[t]
    \centering
    \includegraphics[width=0.36\textwidth]{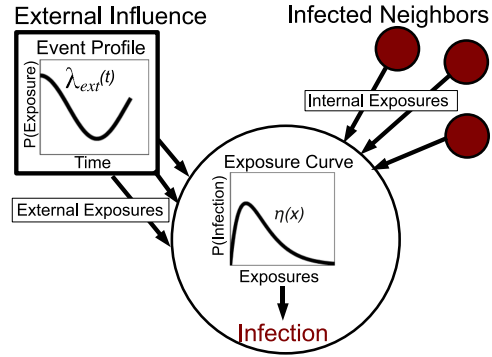}
\vspace{-.3cm}
\caption{Our model of external influence.  A node (denoted by a big circle) is exposed to information through an external source (governed by external activity $\lambda_{ext}(t)$) and by already infected neighbors (governed by the internal hazard function $\lambda_{int}(t)$).  With each new exposure $x$, the probability of infection changes according to the exposure curve $\eta(x)$. 
We infer both the external activity $\lambda_{ext}(t)$, as well as the exposure curve $\eta(x)$.}
\label{fig:flow_chart}
\vspace{-.6cm}
\end{figure}

Networks represent a fundamental medium for the emergence and diffusion of information~\cite{rogers95diffusion}. For example, we often think of information, a rumor, or a piece of content as being passed over the edges of the underlying social network~\cite{jure07cascades,watts02cascades}. This way information spreads over the edges of the network like an epidemic~\cite{hethcote00diseases}. However, due to the emergence of mass media, like newspapers, TV stations and online news sites, the information not only reaches us through the links of our social networks but also through the influence of exogenous out-of-network sources~\cite{bennett-news-illusion}. From the early stages of research on news media and, more generally, information diffusion, there has been the tension between global effects from the mass media and local effects carried by social structure~\cite{lazarsfeld-peoples-choice}.


Traditionally, it was hard to capture and study the effects of mass media and social networks simultaneously~\cite{katz57twostep}. However, the Web, blogs and social media changed the traditional picture of the dichotomy between the local effects carried by the links of social networks and the global influence from the mass media. Today, mass media as well as the social networks both exist in the same Web ``ecosystem,'' which means that it is possible to collect massive online social media data and at the same time capture the effects of mass media as well as the influence arising from the social networks~\cite{jure09memes}. This allows us to study processes of information diffusion and emergence in much finer detail than ever before. 


In this paper, we ask the question ``How does information transmitted by the mass media interact with the personal influence arising from social networks?''  Based on the {\em complete} one month Twitter data we study ways in which information reaches the nodes of the Twitter network. We analyze over 3 {\em billion} tweets to discover mechanisms by which information appears and spreads through the Twitter network. In particular, we contrast two main ways by which information {\em emerges} at the nodes of the network: {\em diffusion} over the edges of the network, and {\em external influence}, when information ``jumps'' across the network and appears at a seemingly random node. For example, when information appears at a node that has no connections to nodes that have previously mentioned the information, the emergence of information at that node can only be explained by the influence of some unobserved exogenous source. However, when information appears at a node with a neighbor that already tweeted it, then it is not clear whether the node tweeted the information due to neighbor's influence or due to the influence of the exogenous source. Thus, the effect of internal and external infections get confounded~\cite{aral09contagion} and the goal of the paper is to develop models that will allow us to separate the influence transmitted by social networks from the influence of the exogenous source(s).


\xhdr{Effects of external influence}
%
On Twitter, users often post links to various webpages --- most often these are links to news articles, blog posts, funny videos or pictures.
Generally there are two fundamental ways how users learn about these URLs and tweet them. One would be due to the exogenous out-of-the-network effects. For example, one can imagine a scenario, where one checks news on CNN.com, finds an interesting article and then posts a tweet with a URL to the article. In this case CNN is the ``external influence'' that caused that URL to emerge onto a particular Twitter user. At contrast, users can also come across URLs by seeing them posted by other users that they follow.  This type of user-to-user exposure is what we refer to as ``internal influence,'' or diffusion.  We find that both external and internal influence play significant role in the emergence of URLs in the Twitter network. 

\xhdr{Modeling the external influence}
In order to accurately model the emergence of content in Twitter we need to consider the activity of the invisible out-of-network sources that also transmit information to the nodes of the Twitter network (via channels, like TV, newspapers, etc.). We present a probabilistic generative model of information emergence in networks, in which information can reach a node via the links of the social network or through the influence of the external source. Developing such a model is important.
For example, we simulated a purely non-diffusive process that picks nodes of the Twitter network at random and `infects' them. After such process infects 10\% of the nodes, about 30\% of infections (falsely) appear to be a result of diffusion, \ie, the random process picks a node that has (simply by chance) an already infected neighbor. Thus, instead of estimating the amount of internal influence at 0\%, naive estimate would be 30\%. Due to the confounding of diffusion and external influence, we aim to separate the two factors.

In our model (Figure \ref{fig:flow_chart}) we distinguish between {\em exposures} and {\em infections}~\cite{romero11twitter}. An {\em exposure event} occurs when a node gets exposed to information $I$, and an {\em infection event} occurs when a node posts a tweet with information $I$. Exposures to information lead to an infection. A node can get exposed to information in two different ways. First, a node $U$ gets exposed to or becomes aware of information $I$ whenever one of his neighbors in the social network posts a tweet containing $I$ (we call this an {\em internal exposure}). The second way $U$ can be exposed to $I$ is through the activity of the external source (we refer to this as {\em external exposure}). We refer to the volume of external exposures over time as the {\em event profile}. In order to establish the connection between exposures and infections, we define the notion of the {\em exposure curve} that maps the number of times node $U$ has been exposed to $I$ into the probability of $U$ getting infected~\cite{romero11twitter}. Distinguishing between exposures and infections, and explicitly modeling the exposure curve allows us to capture rich effects. For example, during the diffusion of a news story, the story may become stale and less relevant each time a user sees it, so the probability of infection would decrease with each exposure.  On the other hand, exposures to a story about new technology may have the opposite effect; with each exposure the user learns more about the technology so the probability of infection would increase. Exposure curves allow us to model such diverse behaviors that our model is able to accurately estimate from the data.

Furthermore, we also develop an efficient parameter estimation technique. We are given a network and a set of node infection times. We then infer the event profile, which quantifies the number of exposures generated by the external source over time. We also infer the exposure curve that models the probability of infection as a function of the number of exposures of a node. Our model accurately distinguishes external influence from network diffusion.

We experiment with our model on Twitter and find that we can accurately detect the occurrence external out-of-network events, and the exposure curve inferred from our model is often 50\% more accurate than baseline methods. We find even though we are studying processes intrinsic to the Twitter network, only about 71\% of the content that appears in Twitter can be attributed to the diffusion through the edges of the Twitter network. We fit our model to 18,186 different URL's that have appeared across Twitter users, and we use the inferred parameters of the model to provide insights into the mechanics of the emergence of these URLs. Moreover, we also perform per topic analysis and find that topics, like Politics and Sports, are most heavily driven by the external sources, while Entertainment and Technology are driven internally with only $\sim18$\% of exposures being external.


\section{Related Work}
\label{sec:related}
%

Work on the diffusion of innovations~\cite{rogers95diffusion} provides a conceptual framework to study the emergence of information in networks. Conceptually, we think of an (often implicit) network where each node is either \emph{active} (infected, influenced) or \emph{inactive}, and active nodes can then spread the \emph{contagion} (information, disease) along the edges of the underlying network. A rich set of models has been developed that all try to describe different mechanisms by which the contagion spreads from the infected to an uninfected node \cite{centola-complex-contagions,cosley10influence,jure07cascades,romero11twitter,watts02cascades}. However, nearly all models only focus on the diffusive part of the contagion adoption process, while neglecting the external influence. In this regard our work introduces an important dimension to the diffusion of innovations framework, where we {\em explicitly} model the activity and influence of the external source.

External influence in networks has been considered in the case of the popularity of YouTube videos~\cite{crane08response}. Authors considered a simple model of information diffusion on an implicit completely connected network and argued that since some videos became popular quicker than their model predicted, the additional popularity must have been a result of external influence. Our approach differs significantly: We directly consider the network and the effect of node-to-node interactions, explicitly infer the activity of external source over time and use a much more realistic model of information adoption that distinguishes between exposures to and the adoption of information. Our model builds on the notion of exposure curves which was proposed and studied by Romero et al.~\cite{romero11twitter}. Recently, it was also argued \cite{steeg11epidem} that it is the shape of exposure curves that stops the information from spreading. We make a step forward by providing an inference method that infers the shape of such exposure curves. Simulations show that our method much more accurately infers the exposure curves than the methods previously proposed~\cite{romero11twitter,steeg11epidem}.


\section{Proposed Model}
\label{sec:proposed}

Here, we develop in detail our novel information diffusion model that incorporates both the spread of information from node to node along edges in the network as well as the external influences acting on the network. Additionally, our model reconciles the gap between a stream of exposures arriving in continuous time and a series of discrete decisions leading to infection.

We refer to the amount of influence external sources have on the network as a function of time as the {\em event profile}.  It is proportional to the probability of any node receiving an external exposure at a particular time. We use the term {\em contagion} to refer to a particular piece of information emerging in the Twitter network and we say a node is {\em infected} with a particular contagion when she first mentions/tweets the contagion. We model contagions as independent of each other, which means we consider them one by one.

We illustrate our model in a node-centric context in Fig. \ref{fig:flow_chart}.  Assume a single contagion (i.e., a piece of information). As time progresses, a node receives a stream of varying intensity of external exposures, governed by the event profile $\lambda_{ext}(t)$.  Additionally, its neighbors in the network also become infected by the contagion, and each infected neighbor generates an internal exposure.  Each exposure has a chance of infecting the node, but with the arrival of each exposure, the probability of infection changes according to the exposure curve $\eta(x)$.  Eventually, either the arrival of exposures will cease, or the node will become infected and then expose to its neighbors. Our goal is to infer the number of exposures generated by the external source over time, as well as the shape of the exposure curve $\eta(x)$ that governs the probability of node's infection.

\begin{table*}
\centering
\begin{tabular}{|c|c|c|c|}
\hline
Symbol & Name & Description & Technical Definition \\
\hline
$\lambda_{ext}(t)$ & The Event profile & \begin{minipage}{.275\linewidth}\vspace{.15cm}\centering Proportional to the probability of any node receiving an exposure at time $t$. \vspace{.15cm}\end{minipage}& $\lambda_{ext}(t)\,dt=P[\mbox{ node exposed } \in [t, t+dt)]$\\
 \hline
 $\lambda_{int}(t)$ & \begin{minipage}{.15\linewidth}\centering Internal Hazard Function \end{minipage}  & 
 \begin{minipage}{.275\linewidth}\vspace{.15cm}\centering  Governs the random amount  of time it takes an infected node to expose its neighbors\vspace{.15cm}\end{minipage} &

 \begin{minipage}{.375\linewidth}\centering
$
 	\lambda_{int}(t) \, dt = $ \\$
	P\left( i \mbox{ exposes } j \in [t, t+dt] \right|  i  \mbox{ hasn't exposed } j \mbox{ yet}) 
 $
 \end{minipage}\\

 \hline
 $\eta(x)$ &\begin{minipage}{.15\linewidth}\centering The Exposure Curve \\ (parameters $\rho_1$, $\rho_2$)  \end{minipage}   &\begin{minipage}{.275\linewidth}\vspace{.15cm}\centering  Determines how the probability of infection changes with each exposure. \vspace{.05cm}\end{minipage} & $\eta(x) = P(  \mbox{ infected right after } x^{th} \mbox{ exposure} )$ \\

\hline
$P_{exp}^{(i)}(n;t)$ & The Exposure Distribution &\begin{minipage}{.275\linewidth}\vspace{.15cm}\centering  The probability that node $i$ has received $n$ exposures by time $t$\vspace{.15cm}\end{minipage} & 

 \begin{minipage}{.375\linewidth}\centering \vspace{.1cm}
$P_{exp}^{(i)} = { t / \Delta t \choose {n}} \left(\frac{\Lambda^{(i)}_{int}(t) + \Lambda_{ext}(t)}{t}\cdot \Delta t\right)^{n} $\\
$\times \left( 1 - \frac{\Lambda^{(i)}_{int}(t) + \Lambda_{ext}(t)}{t}\cdot \Delta t \right)^{t / \Delta t - n}$
\vspace{.1cm}
\end{minipage}\\

\hline
$\tau_i$	& Infection time  & The infection time of node $i$& \\
\hline
\end{tabular}
\caption{Definition of symbols used in the model.}
\label{tab:sym_def}
\end{table*}

\xhdr{Modeling the internal exposures}
Consider a single contagion. In our model, an internal exposure occurs when a neighbor of a node becomes infected, and then an exposure is transmitted after a random interval of time.  Imagine a real world scenario in which the social network is the Twitter network and the contagion spreading across the network is a particular URL.  If a neighbor writes a tweet involving a particular URL then a user sees their neighbor's tweet, then and only then has the internal exposure propagated along the edge.
An infected node will expose each of its outgoing neighbors exactly once, and the time it takes for each exposure to occur is sampled from some distribution universal to all edges in the network.  Therefore, a hazard function \cite{johnson11hazard} is appropriate to model this process. Hazard functions were originally developed in actuary sciences, and they describe a distribution of the length of time it takes for an event to occur. Recently, \cite{rodriguez11} used hazard functions as a basis for disease propagation in continuous time across social networks.  They are extremely effective at modeling discrete events that happen over continuous time.  In this respect hazard functions represent a principled way of occurrence of discrete events (\ie, exposures) as a function of continuous time.

Specifically, let $\lambda_{int}$ be the \textit{internal hazard function}, where \vspace{-.1cm}
\begin{align*}
\lambda_{int}(t) \, dt \equiv P\left( i \mbox{ exposes } j \in [t, t+dt) \right|  i  \mbox{ hasn't exposed } j \mbox{ yet}) \vspace{-.3cm}
\end{align*}
for any neighboring nodes $i$ and $j$, where $t$ is the amount of time that has passed since node $i$ was infected. In our context, $\lambda_{int}$ effectively models how long it takes a node to notice one of its neighbors  becoming infected.  It is a function of the frequency with which nodes check-up on each other.  For the Twitter network, each time a user logs-in they are updated on all of their neighbors. 

The expected number of internal exposures a node $i$ has received by time $t$, which we will define as $\Lambda_{int}^{(i)}(t)$, is the sum of the cumulative distribution functions of exposures propagating along each of the node's inbound edges and can be derived as follows:
\begin{align}
\Lambda_{int}^{(i)}(t) &= \sum_{j; j\mbox{ is \textit{i}'s inf. neighbor } } P( j \mbox{ exposed } i \mbox{ before } t)\\
&=\sum_{j; j\mbox{ is \textit{i}'s inf. neighbor } }  \left[ 1 -  \exp \left( -\int_{\tau_j}^t \lambda_{int}(s - \tau_j )ds\right) \right]
\label{eqn:L_int}
\end{align}
where $\tau_j$ is the infection time of node $j$.

\xhdr{Modeling the external exposures}
The second source of exposures to a particular single contagion for nodes in the network comes from the external source acting on the network.  The fundamental property of the problem we are trying to solve is that the external source cannot be observed.  
The source varies in intensity over time, and this function is called the \textit{event profile}, which we designate as $\lambda_{ext}(t)$.  Specifically,
\vspace{-.1cm}
\begin{align*}
    \lambda_{ext}(t)\,dt \equiv P( i \mbox{ receives exposure} \in [t, t+dt))\vspace{-.3cm}
\end{align*}
for any node $i$, where $t$ represents the amount of time since the contagion first appeared in the network.  A couple of things should be noted here. First, all nodes have the same probability of receiving an external exposure for any point in time. Second, $\lambda_{ext}$ is not conditioned upon the node not already having received an external exposure.  This means that any node can receive an arbitrary number of external exposures.  
We call $\lambda_{ext}$ the event profile because it describes an actual real world event that caused the information to arrive in the network and start spreading.  As the event progresses over time, event's efficacy in the network changes.  For example, if our contagion is civil unrest in Libya, then every time their ruler Gaddafi gives a speech or the rebels win a battle we would expect a spike in the intensity of the external source and thus the even profile $\lambda_{ext}$.  As time passes without any new developments or as the event's relevancy fades, we expect $\lambda_{ext}$ decrease to 0. However, every time there is a new development we expect a spike in the external event profile $\lambda_{ext}$. We will infer $\lambda_{ext}$ non-parametrically, so we can quantify the relevancy of any event over its lifespan.

In order to derive the distribution of exposures a node receives over time as a function of time, we model the arrival of exposures as a binomial distribution.  Consider we were to take the entire continuous time interval of the lifetime of the contagion and break it down into smaller but finite time intervals. Then whether an exposure occurred during each such subinterval is a Bernoulli random variable (exposure vs. no exposure) with its own probability.  Therefore, the total number of exposures received in a time interval is a sum of Bernoulli random variables, just as a binomial random variable is a sum of Bernoulli random variables.  Let's say that $\lambda_{ext}$ is constant for all time and that time is discretized into finite intervals of length $\Delta t$.    Then the probability that $n$ external exposures have been received after $T$ time intervals is exactly a binomial distribution:
\begin{align*}
P_{exp}(n; T \cdot \Delta t) = {T \choose {n}} \left(  \lambda_{ext}\cdot \Delta t \right)^{n}\cdot \left( 1 - \lambda_{ext}\cdot \Delta t \right)^{T - n}.
\end{align*}
Set $t=T\cdot \Delta t$.  If we take the limit of as $\Delta t \rightarrow 0$ and $T\rightarrow \infty$ such that $t$ does not change, then this probability approaches
\begin{align*}
P_{exp}(n; t) = {t / dt \choose {n}} \left(  \lambda_{ext}\cdot dt \right)^{n}\cdot \left( 1 - \lambda_{ext}\cdot dt \right)^{t/dt - n}.
\end{align*}
To relax the constraint that $\lambda_{ext}$ is constant, we use the average of $\lambda_{ext}(t)$ over $t$:
\begin{align*}
P_{exp}^{(i)}(n; t) &\approx{ t / dt \choose {n}} \left( \frac{\Lambda_{ext}(t)}{t}\cdot dt \right)^{n}
\cdot \left( 1 - \frac{ \Lambda_{ext}(t)}{t}\cdot dt \right)^{t / dt - n}.
\end{align*}
where $\Lambda_{ext}(t)\equiv \int_0^t\lambda_{ext}(s)ds$.  Finally, users are receiving both external and internal exposures at the same time, so if we need take into account both processes.  This would imply taking the convolution of the two probabilities, which would be computationally infeasible.  Instead, we use the average of $\lambda_{ext}(t) + \lambda^{(i)}_{int}(t)$:

\begin{align}
P_{exp}^{(i)}(n; t) &\approx { t / dt \choose {n}} \left( \frac{\Lambda^{(i)}_{int}(t) + \Lambda_{ext}(t)}{t}\cdot dt \right)^{n} \\
&\times \left( 1 - \frac{\Lambda^{(i)}_{int}(t) + \Lambda_{ext}(t)}{t}\cdot dt \right)^{t / dt - n}
\label{eqn:xi_i}
\end{align}

Effectively, we approximated the flux of exposures as constant in time such that each interval of time has an equal probability of an exposure arriving, so the sum of the events is a standard binomial random variable.

\begin{figure}
    \includegraphics[width=.19\textwidth]{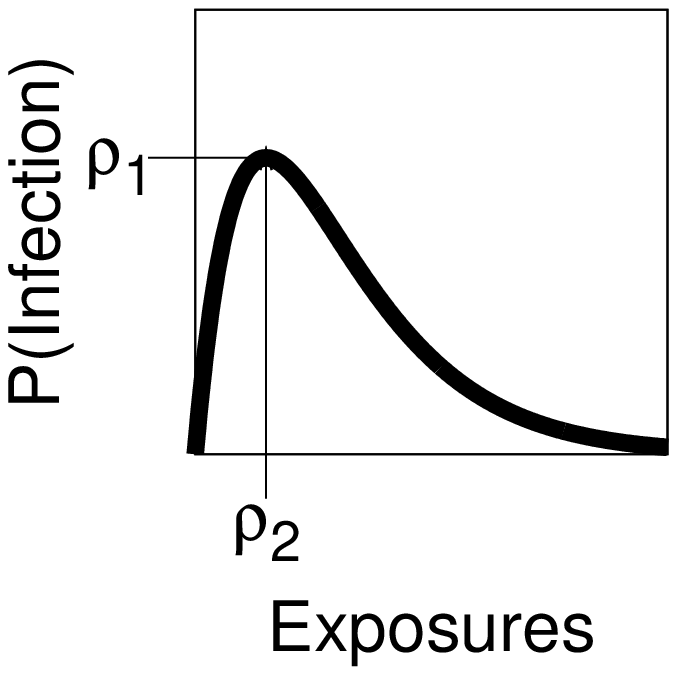}\hspace{-.9cm}
    \includegraphics[width=.19\textwidth]{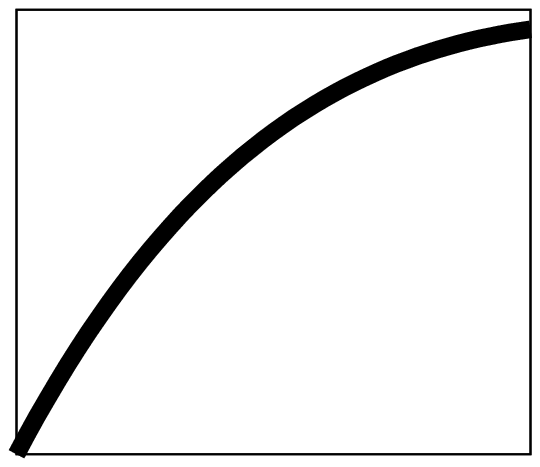}\hspace{-.9cm}
    \includegraphics[width=.19\textwidth]{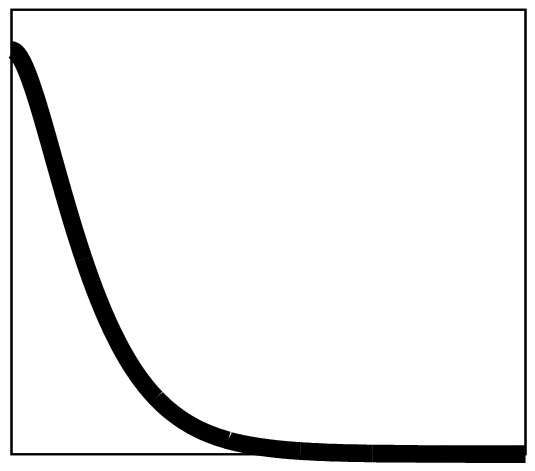}\hspace{-.9cm}
    \vspace{-.3cm}
\caption{Example exposure curves $\eta(x)$, where $\eta(x)$ is the probability of a node becoming infected upon its $x^{th}$ exposure to the contagion.  The parameters of $\eta(x)$ are $\rho_1$ and $\rho_2$.}
    \vspace{-.3cm}
\label{fig:sample_eta}
\end{figure}

\xhdr{Modeling the exposure curve}
We model the exposure curve as a parameterized equation.  Recall that the exposure curve describes the probability of infection as a function of the number of exposures received.  More specifically, if $x$ is the current number of exposures the node has received and $\eta
(x)$ is the exposure curve, then
\begin{align*}
\eta(x) \equiv P(  \mbox{node } i \mbox{ is infected immediately after } x^{th} \mbox{ exposure} ).
\end{align*}
We choose to parameterize $\eta(x)$ as
\begin{align*}
\eta(x) &= \frac{ \rho_1}{\rho_2}\cdot x \cdot \exp \left( 1 - \frac{x}{\rho_2} \right).
\end{align*}
where $\rho_1\in (0,1]$ and $\rho_2 > 0$.  Parameterizing $\eta(x)$ in this manor allows for several desirable properties.  First, $\eta(0)=0$ so it is impossible to become infected by a contagion before being exposed to it.  Secondly, this function is unimodal with an exponential tail, so there is a critical mass of exposures when the contagion is most infectious followed by decay brought on by the idea becoming overexposed/tiresome.  Lastly, and most importantly, $\rho_1$ and $\rho_2$ have important conceptual meanings: $\rho_1 = \max_x \eta(x)$ and $\rho_2 =\mbox{arg}\max_x \eta(x).$ Because of this, we can think of $\rho_1$ as a general measure of how infectious a contagion is in the network and $\rho_2$ as a measure of the contagion's enduring relevancy.  Fig. ~\ref{fig:sample_eta} shows several different forms of $\eta(x)$.  This parameterization is expressive, but any other parameterization for $\eta(x)$ is also valid.  For the remainder of the paper, we will discuss the model in the context of the $\eta(x)$ parameterization presented above.

\xhdr{From exposures to infections}
In order to fit the parameters of the model to observed data, we must now construct the probability functions to describe the model.  With the equations given above, building the distribution of the infection time of a node $i$ can be done as follows.  Let $F^{(i)}(t)\equiv P(\tau_i \leq t)$ be the probability that node $i$ has been infected by time $t$, where $\tau_i$ is the infection time of node $i$.  Making use of the quantity $P_{exp}^{(i)}(n;t)$,

\begin{align}
F^{(i)}(t)&=\sum_{n=1}^{\infty}P[i \mbox{ has } n \mbox{ exp. }]
\times P[ i \mbox{ inf. }| i \mbox{ has $n$ exp. }]  \\
&=\sum_{n=1}^{\infty} P_{exp}^{(i)}(n;t) \times \left[ 1 - \prod_{k=1}^n\left[ 1 - \eta(k)\right] \right].
\label{eqn:cum_dist}
\end{align}

While $F^{(i)}(t)$ is analogous to the cumulative distribution function of the infection probability, it is important to note that it is not {\em actually} a distribution;  $\lim_{t\rightarrow \infty}F(t) < 1$ as a result of $\lim_{x\rightarrow \infty} \eta(x)=0$.  This is ideal because it implies that there is a non-zero chance that a node will never become infected, as should be the case.


\subsection{Inferring the model parameters}

Next we develop a method of inferring the model parameters for a given network and the tract of a single contagion. We fit the model to each contagion separately.  We are given the network and the infection times for each node that got infected with the contagion under consideration. We then need to infer the event profile $\lambda_{ext}(t)$ for all $t$ at which at least one node was infected, and parameters of $\eta(x)$, $\rho_1$ and $\rho_2$, of the exposure curve. In all, the number of parameters we are inferring is the number of {\em unique} node infection times plus the two parameters of $\eta(x)$.  Our general strategy is to alternate back and forth from inferring $\lambda_{ext}(t)$ to inferring $\eta(x)$, assuming we known one for certain while we infer the other, until both functions converge.  Below, we first demonstrate how to infer the event profile when the exposure curve is known.  Then, we show how to infer the exposure curve with a known event profile.  Finally, we combine the two steps into a single algorithm.

\xhdr{Inferring the event profile}
The following outlines a fast and robust method for inferring $\lambda_{ext}(t)$, given $\eta(x)$. Let $S(t)$ be the number of nodes that are uninfected (by the contagion currently under consideration) at time $t$.  $S(t)$ is a random variable whose expectation value is dependent on $\lambda_{ext}(t)$, $\eta(x)$, and the underlying network.  The networks which we are interested in are sufficiently large, so the quantity $S(t) - E\left[ S(t)\right]$ is usually very small in magnitude.  This provides us with a very straight-forward method for inferring $\Lambda_{ext}(t) \equiv \int_0^{t}\lambda_{ext}(s)ds$.  Let $t_k$ be the $k^{th}$ time at which at least one node was infected, then define $\Lambda_k$ as $\Lambda_{ext}(t_k)$.  To calculate $S(t)$,
\vspace{-.3cm}
\begin{align}
S(t_k) &= \sum_{i=1}^N P(\mbox{ node $i$ not infected by time $t$})\\
       &= \sum_{i=1}^N \sum_{n=1}^\infty P_{exp}^{(i)}(n;t_k) \prod_{k=1}^n \left[1 - \eta(k) \right]\\
       &\approx \sum_{i=1}^N \sum_{n=1}^\infty P_{exp}^{(i)}(n;t_k) \exp\left( -\int_{y=0}^n \eta(y)dy \right) \\
       &\approx \sum_i \exp \left( -\int_0^{\Lambda_k + \Lambda^{(i)}_{int}(t_k)}\eta(y)dy \right).
\label{eqn:solve_for_event_profile}
\end{align}
The first approximation comes from treating the number of exposures received by a node at any given time as a continuous real number instead of an integer.  This provides us with a closed-form expression.  The second approximation comes from setting the number of exposures received by each node to be the expected number of exposures.

Since the right-hand side is monotonic (it is strictly decreasing with respect to $\Lambda_k$), we can solve for $\Lambda_k$ using bisection search.  Doing this for all $t_k$ gives us $\Lambda_{ext}(t_k)$ for each possible time, and then we can use finite difference to get $\lambda_{ext}(t_k)$.

Once the event profile has been inferred, we must then update the exposure curve accordingly.

\xhdr{Inferring the exposure curve}
Now, we assume we know $\Lambda_{ext}(t)$ for all $t_k$, and we want to infer the exposure curve $\eta(x)$, specifically its parameters $\rho_1$ and $\rho_2$.  Our strategy in solving for these parameters will be to fix $\rho_2$, and then solve for a $\rho_1$ that maximizes the following approximation to the log-likelihood.  Making use of Eq. \ref{eqn:cum_dist}, we have
\vspace{-.3cm}
\begin{align*}
\mathcal{L}(\eta, \Lambda_{ext},& \lambda_{int}) = \sum_{i\in {\mathcal{I}}} \log\left[ \frac{d[F^{(i)}(t)]}{dt} \right] +\sum_{i\in {\mathcal{I}^c}} \log\left[ F^{(i)}(t) \right] \\
&\approx \sum_{i\in {\mathcal{I}}}\sum_{n=1}^\infty P_{exp}^{(i)}(n;\tau_i)\left[ \log( \eta(n) ) + \sum_{k=1}^{n-1}\log(1 - \eta(k)) \right]  \\
&+\sum_{i\in {\mathcal{I}^c}}\sum_{n=1}^\infty P_{exp}^{(i)}(n;\tau_{max})\cdot \sum_{k=1}^{n}\log(1 - \eta(k))
\end{align*}
where $\mathcal{I}$ is the set of all infected nodes, $\mathcal{I}^c$ is the set of all uninfected nodes, and $\tau_{max}$ is the time of the last observed infection.  The optimal $\rho_1$ satisfies $\frac{\partial \mathcal{L} }{\partial \rho_1 }=0$ so
\begin{align}
0 &= \frac{\left| \mathcal{I} \right| }{\rho_1} + \sum_{i \in \mathcal{I} } \sum_{n=1}^\infty P_{exp}^{(i)}(n;\tau_{max})\cdot \sum_{k=1}^{n-1}\frac{\eta(k)}{\rho_1 \cdot (1 - \eta(k))} \\
&+\sum_{i\in {\mathcal{I}^c}}\sum_{n=1}^\infty P_{exp}^{(i)}(n;\tau_{max})\cdot \sum_{k=1}^{n}\frac{\eta(k)}{\rho_1 \cdot (1 - \eta(k))}.
\label{eqn:solve_for_eta}
\end{align}
The parameter $\rho_1$ can be solved iteratively, using and initial value between 0 and 1.  Because $P_{exp}^{(i)}$ is independent of $\rho_1$, they only need to be calculated once.  This, along with the iterations converging quickly, makes this entire process very fast.

Now, we combine the event profile inference process with the exposure curve inference process to form a single algorithm that infers the entire model.

\begin{figure}[h!]
    \centering
{\large \textbf{Given}}

\begin{tabular}{ c ||| c }
\hline
    \setcounter{subfigure}{0}\subfigure[All Infections]{\includegraphics[scale=.31]{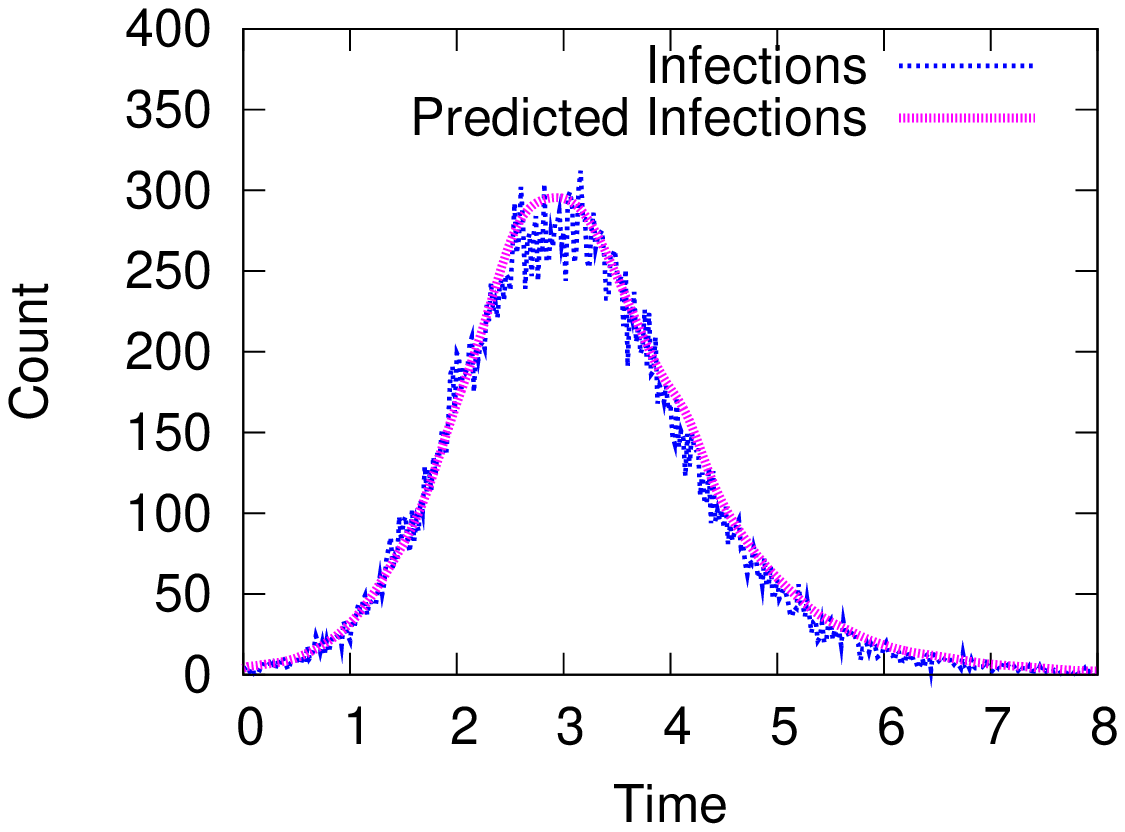}\hspace{-.07cm}\label{fig:syn_dat_1_all} }&
    \setcounter{subfigure}{5}\subfigure[All Infections]{\includegraphics[scale=.31]{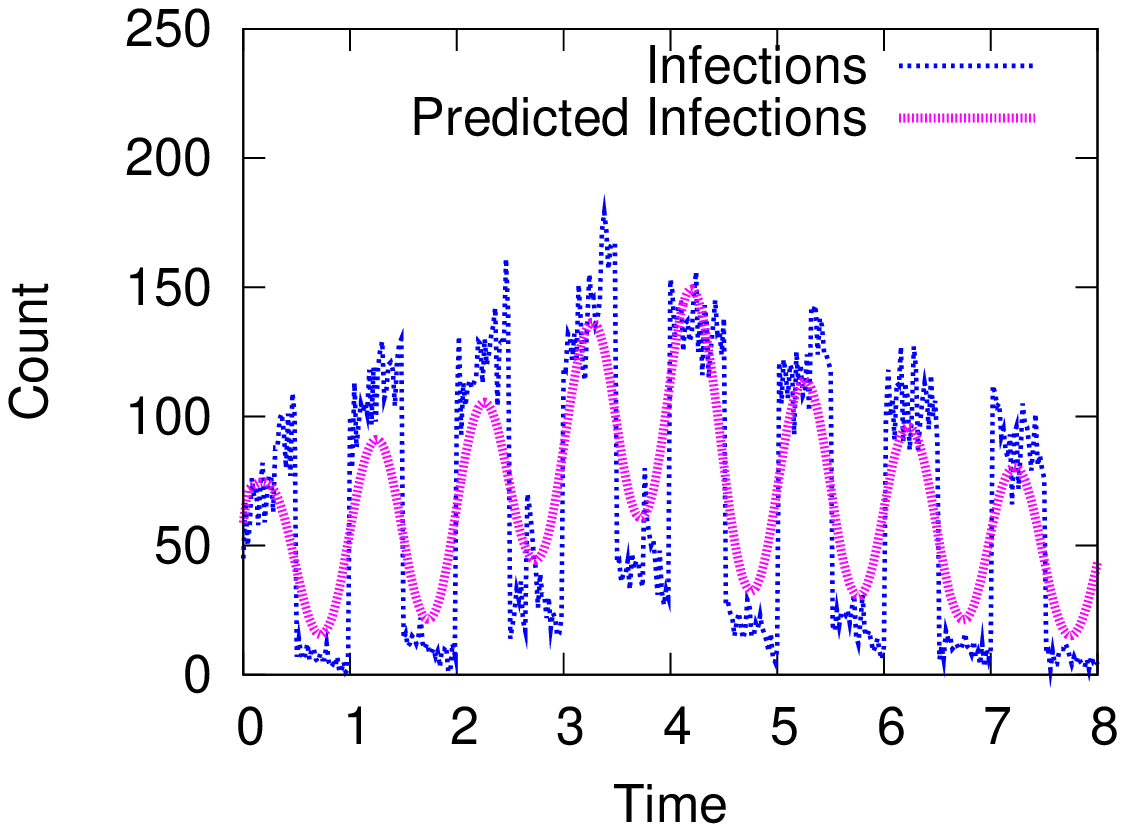}\label{fig:syn_dat_2_all}} \\
\hline
\end{tabular}

\vspace{.1cm}
{\large \textbf{Inferred by the model}}

\begin{tabular}{ c ||| c }
\hline
    \setcounter{subfigure}{1}\subfigure[External Infections]{\includegraphics[scale=.31]{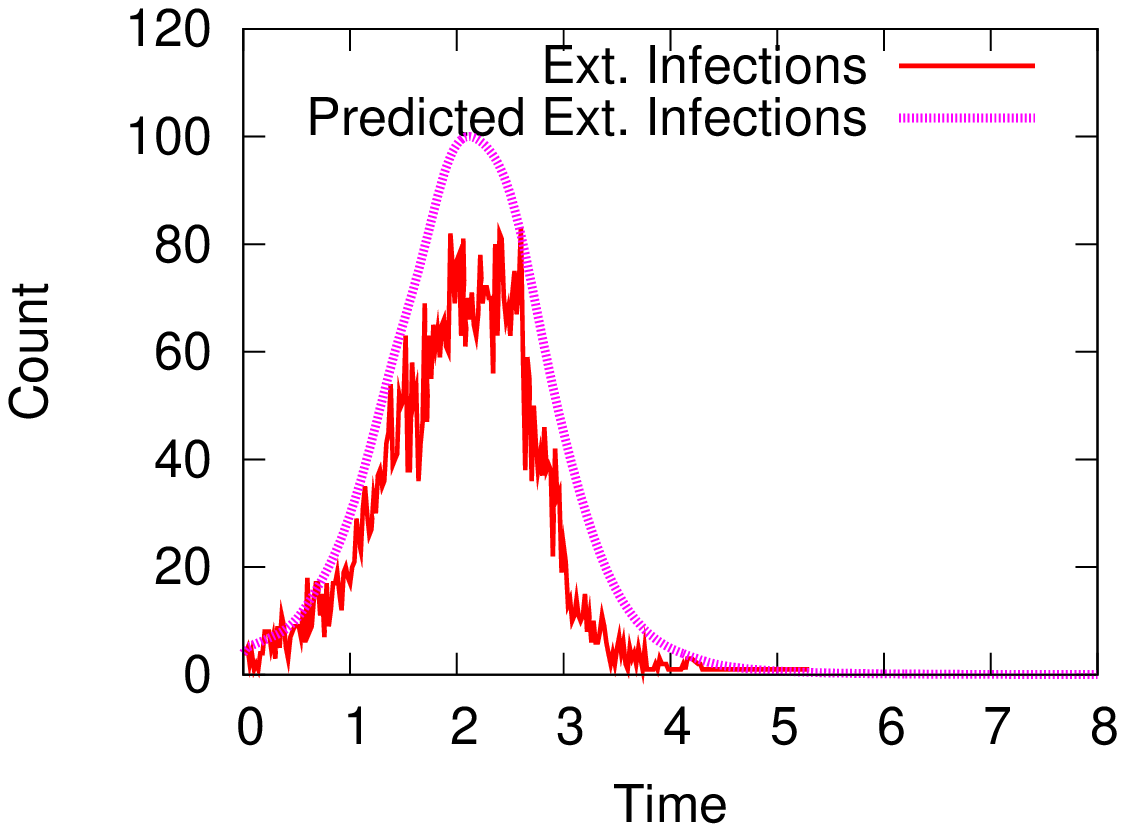}\label{fig:syn_dat_1_ext}} &
    \setcounter{subfigure}{6}\subfigure[External Infections]{\includegraphics[scale=.31]{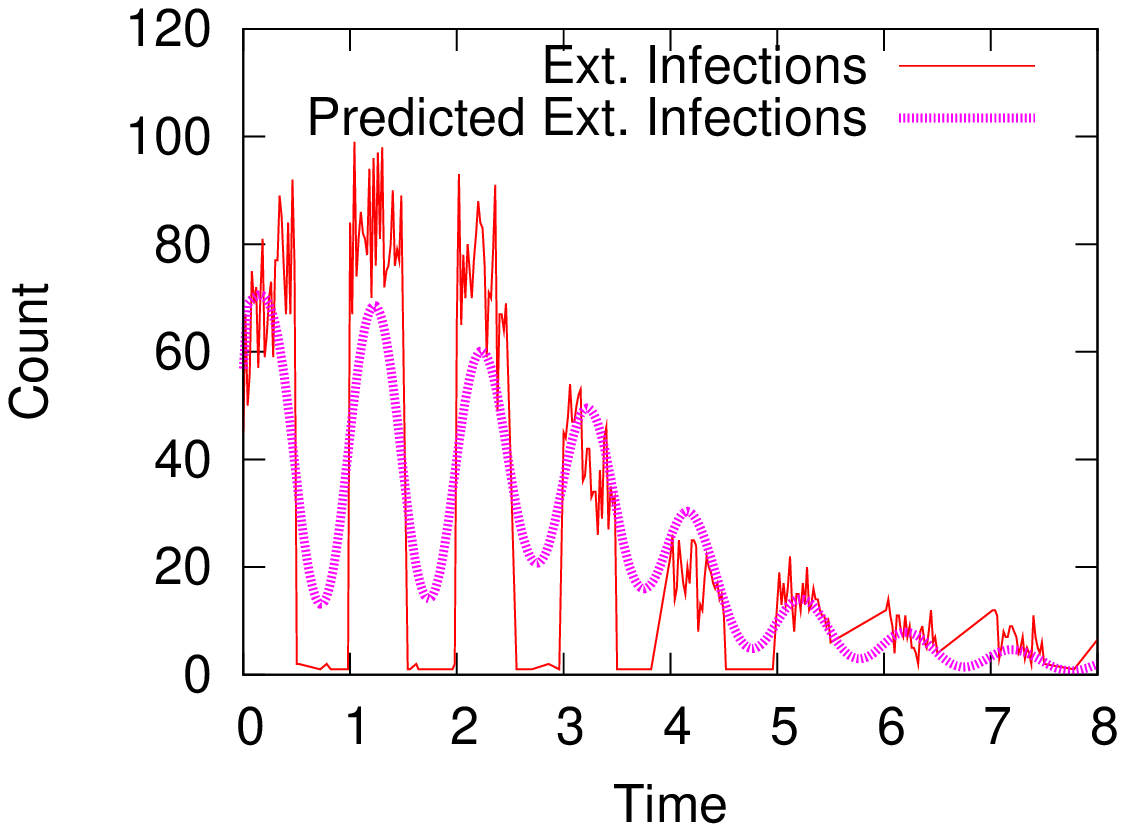}} \\
    \setcounter{subfigure}{2}\subfigure[Internal Infections]{\includegraphics[scale=.31]{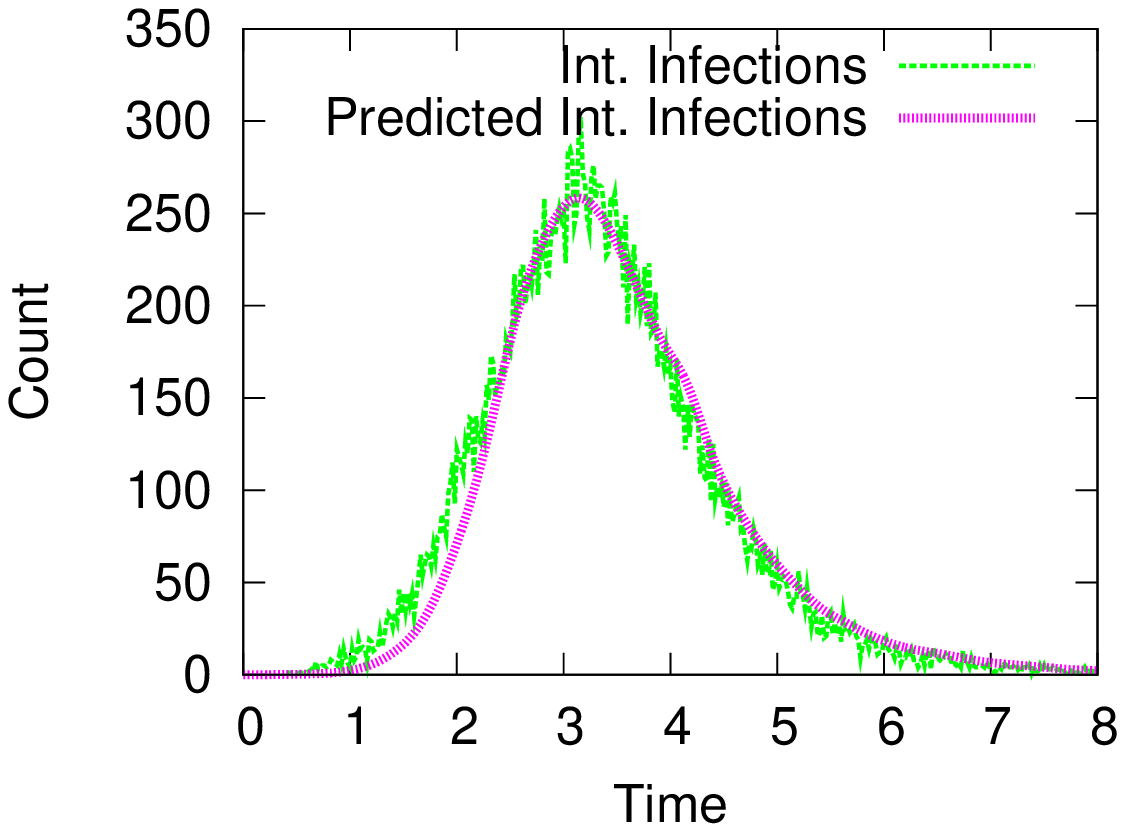}\label{fig:syn_dat_1_int}} &
    \setcounter{subfigure}{7}\subfigure[Internal Infections]{\includegraphics[scale=.31]{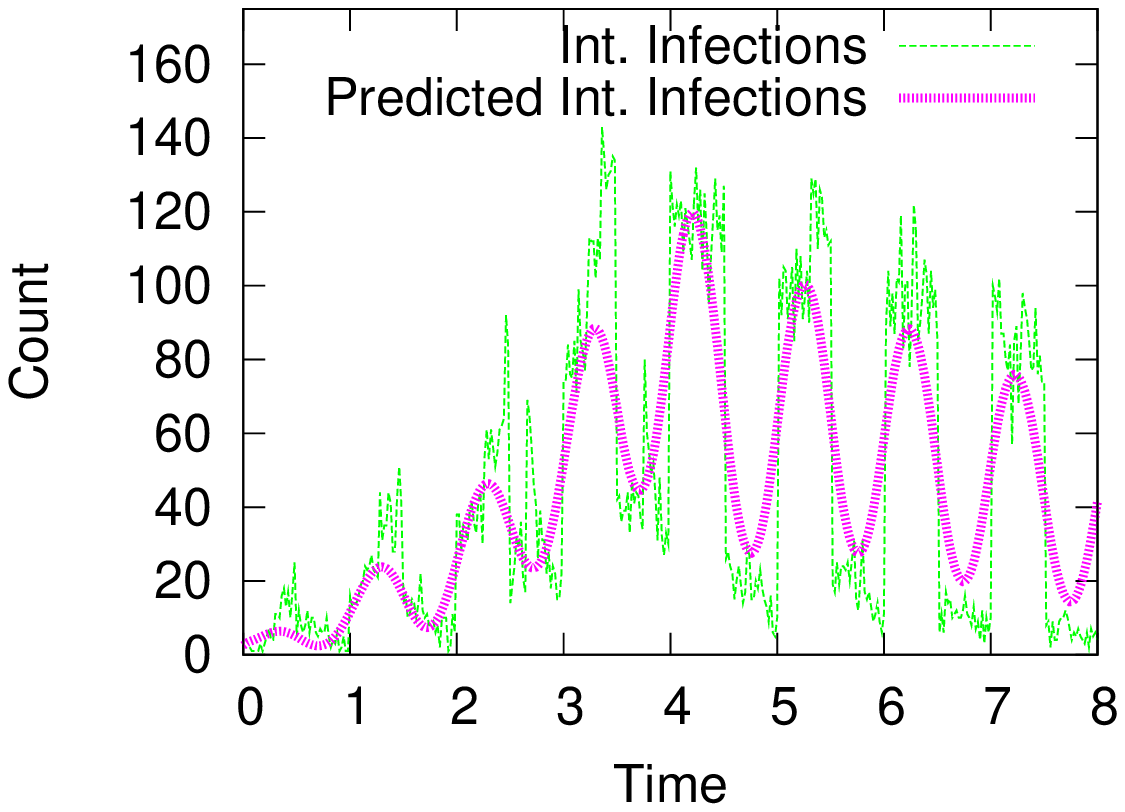}} \\
    \setcounter{subfigure}{3}\subfigure[Event Profile $\lambda_{ext}(t)$]{\includegraphics[scale=.31]{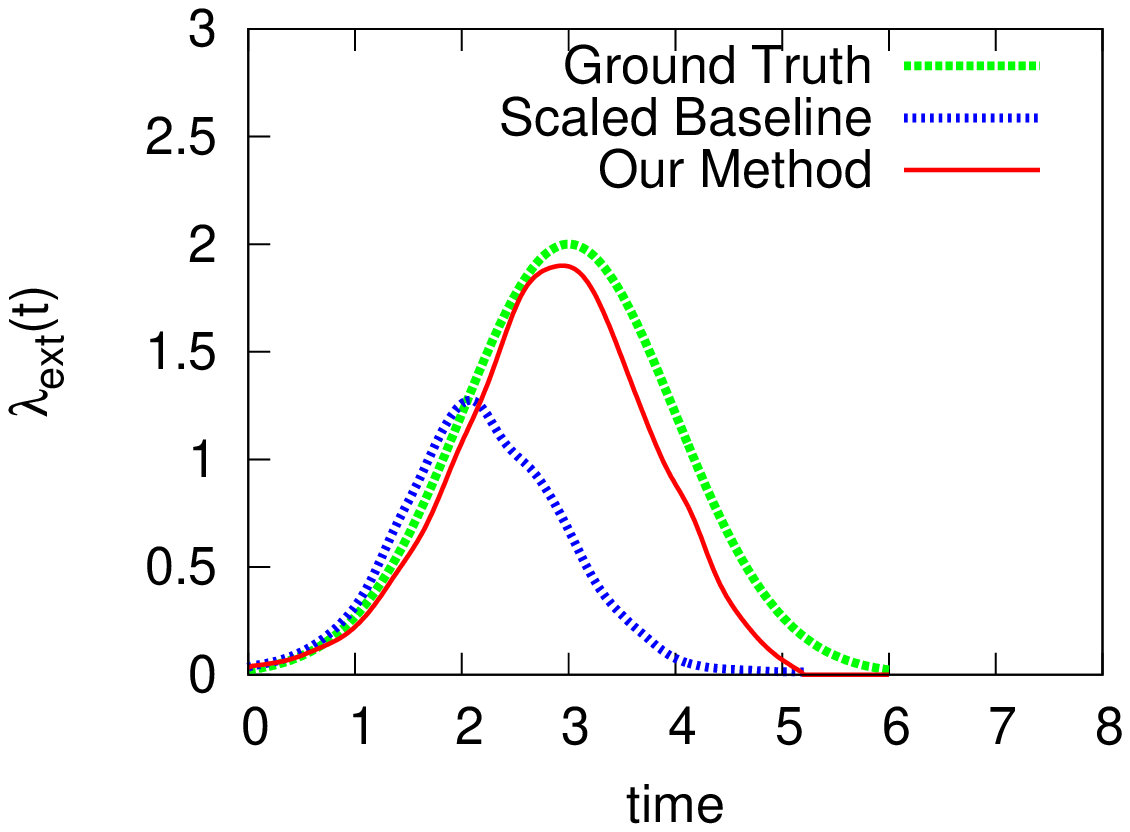}\label{fig:syn_dat_1_lamb}} &
    \setcounter{subfigure}{8}\subfigure[Event Profile $\lambda_{ext}(t)$]{\includegraphics[scale=.31]{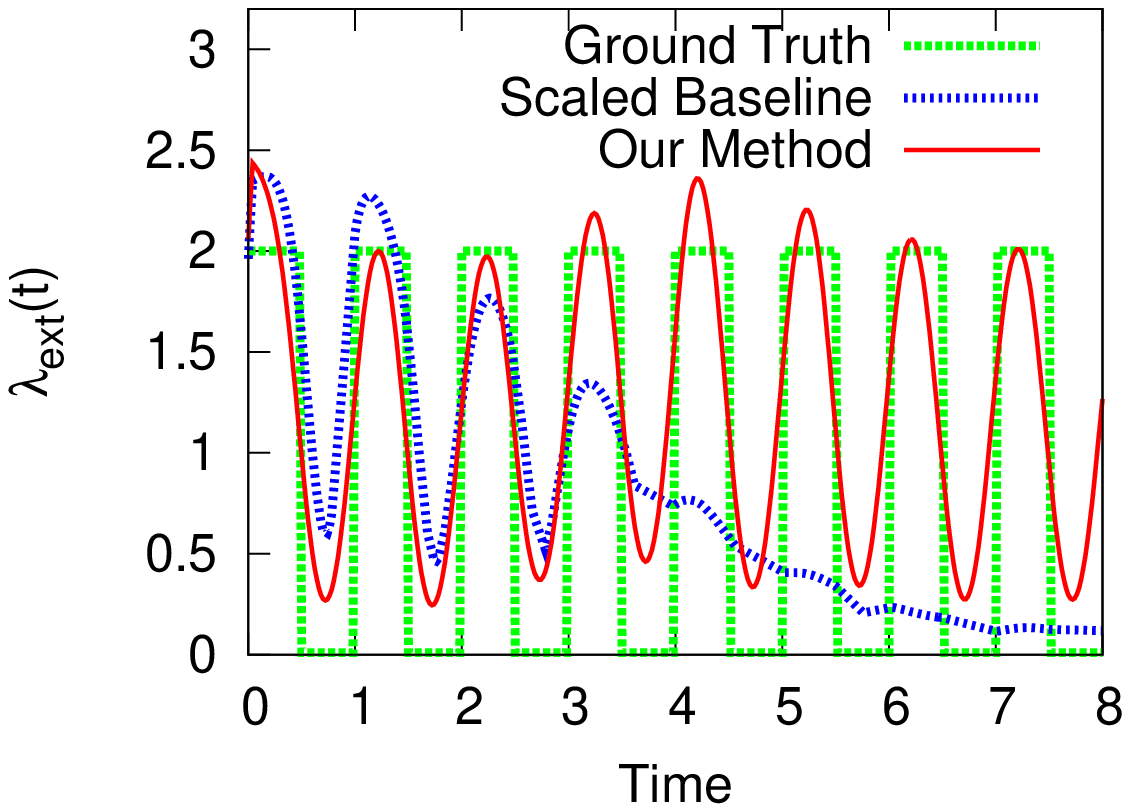}\label{fig:syn_dat_2_lamb}} \\
    \setcounter{subfigure}{4}\subfigure[Exposure Curve $\eta(x)$]{\includegraphics[scale=.31]{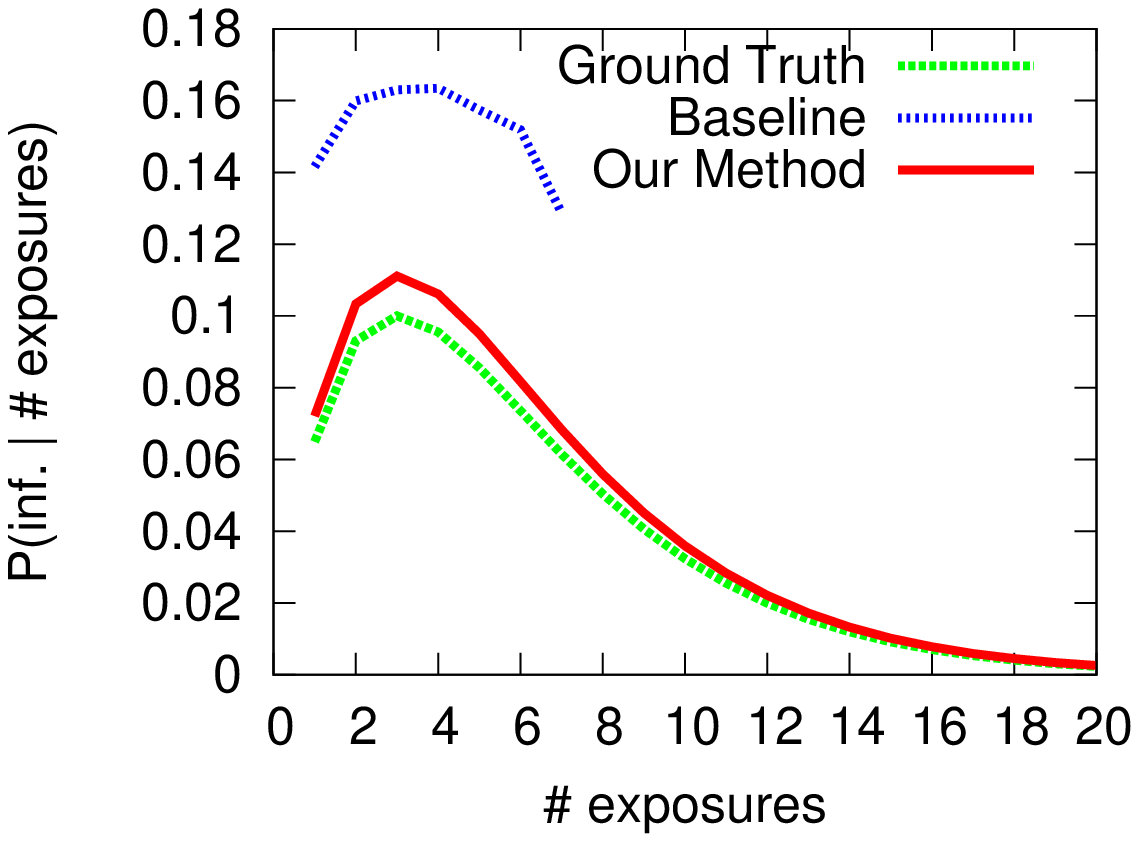}\label{fig:syn_dat_1_eta}} &
    \setcounter{subfigure}{9}\subfigure[Exposure Curve $\eta(x)$]{\includegraphics[scale=.31]{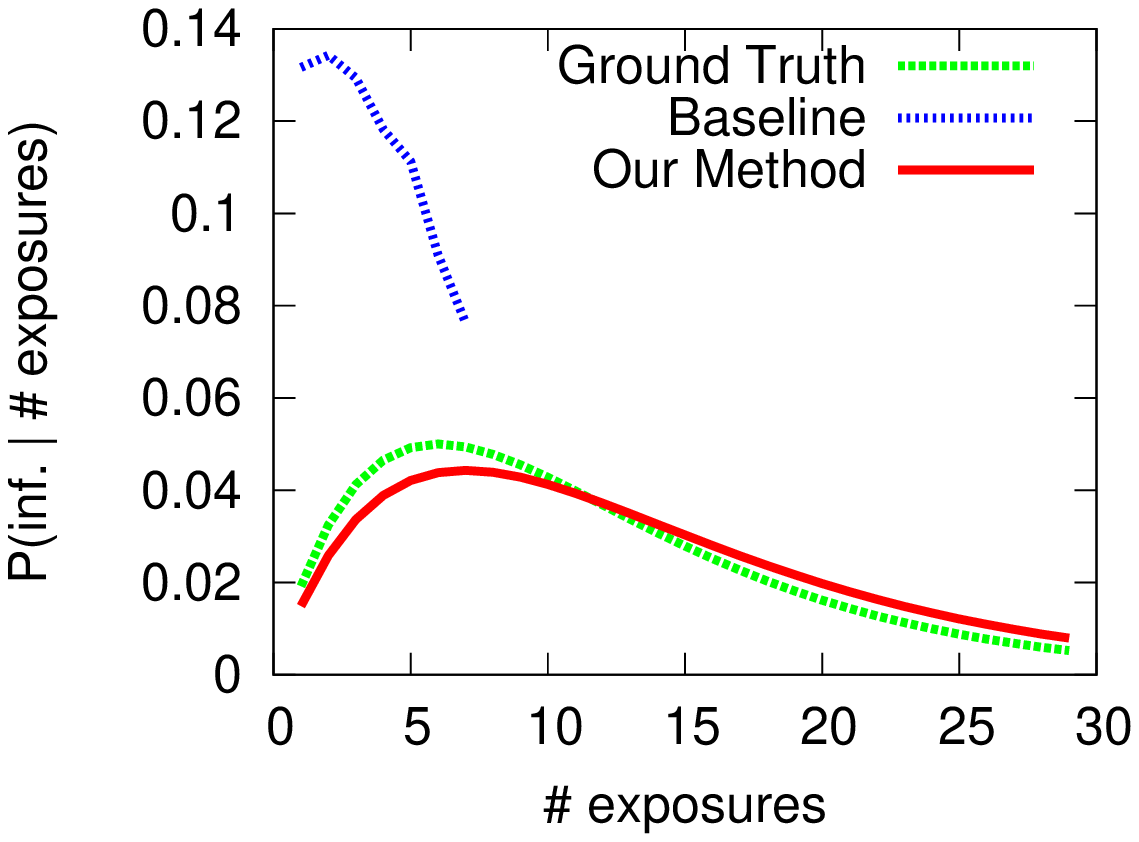}\label{fig:syn_dat_2_eta}} \\
\end{tabular}
\caption{Experiments on synthetic data. (a)-(e) The model fitted to a synthetic contagion on a scale-free network with 75,879 nodes. The internal hazard function is $\lambda_{int}(t)=t$, which induces a Raleigh (unimodal) distribution for the internal exposure propagation time. Given just the number of infections (a) our model is able to infer all of (b)-(e).
(f)-(j) The model fitted to the same network but with the internal hazard function $\lambda_{int}(t)=\frac{1}{t}$, which induces a power law distribution for the internal exposure propagation time.
\vspace{-.5cm}
}
\label{fig:syn_dat}
\end{figure}

\xhdr{Inferring all parameters}
If we use the previously mentioned method to infer $\eta(x)$ using the actual ground-truth $\Lambda_{ext}(t)$, it works extremely well.  In fact, coming up with contrived instances in which it breaks is difficult.  The same thing is true for using the event profile inference method with ground-truth $\eta(x)$.  When neither ground-truth function is known and we have to iterate back and forth between both methods, however, the results are not as stable.  Both functions' inference methods are sensitive to errors in the other function.  Fortunately, all that is needed to correct this is a slight modification.  Simply put, we fix $\rho_2$ to some integer value and then iterate back and forth between the two methods.  Then, $\rho_1$ and $\Lambda_{ext}(t)$ converge to some values dependent on the fixed $\rho_2$, and we calculate the log-likelihood of the resulting inferred functions.  We do this for all reasonable integer values of $\rho_2$, and we choose the one with the optimal log-likelihood.  Algorithm \ref{alg:inference} gives the pseudocode.

\vspace{-.2cm}
\begin{algorithm}
\begin{algorithmic}
\STATE Initialize $\Lambda_{ext}(t)$, $\rho^1_{final}$, $\rho^2_{final}$, $\mathcal{L}_{max}$
\FOR {$\rho_2= 1 \to \rho_{max}$}
      \STATE Initialize $\rho_1$
        \WHILE{not converged}
        \STATE $\rho_1 \gets $ Solution to Eq. \ref{eqn:solve_for_eta} using $\rho_2$, $\Lambda_{ext}(t)$
        \STATE $\Lambda_{ext}(t) \gets $  Solution to Eq. \ref{eqn:solve_for_event_profile} using $\rho_1$, $\rho_2$.
        \ENDWHILE

        $\mathcal{L} \gets $ \textit{Log-Likelihood}$(\Lambda_{ext}(t), \rho_1, \rho_2)$
        \IF {$\mathcal{L} \geq \mathcal{L}_{max}$ }
        	      \STATE $\mathcal{L}_{max} \gets \mathcal{L}$
                \STATE $\rho^1_{final} \gets \rho_1$
                \STATE $\rho^2_{final} \gets \rho_2$
        \ENDIF
\ENDFOR
\STATE $\Lambda_{ext}(t) \gets $  Solution to Eq. \ref{eqn:solve_for_event_profile} using $\rho^1_{final}$, $\rho^2_{final}$.
\end{algorithmic}
\caption{Model Parameter Inference}
\label{alg:inference}
\end{algorithm}
\vspace{-.4cm}
\xhdr{Practical considerations} Since we infer the event profile $\lambda_{ext}(t)$ in non-parametric form, the number of parameters in the model could potentially scale with the time duration of the contagion (we would have to solve for $\lambda_{ext}(t_i)$ for each node's infection time $t_i$).  This can be prevented, however, by predetermining a set of times $\left\{\hat{t}_m \right\}_{m=1}^M$ only at which the event profile will be inferred.  Then, $\lambda_{ext}(t)$ between these set times can be approximated using linear interpolation.  In practice, we used $M=20$, and we set each $\hat{t}_m$ at the time in which $\frac{m}{M}$ of the infections with the contagion have occurred.  Doing this not only makes the runtime constant with respect to the duration of the contagion, it also speeds up the algorithm in general at the price of only a negligible decrease in accuracy.

The algorithm scales linearly with the number of nodes that received at least one exposure.  All nodes that received only external exposures and no internal exposures, however, are effectively identical and can be grouped into a single term for both the event profile inference and the exposure curve inference.  Therefore, in practice the runtime scales linearly with the number of nodes that received {\em at least one internal exposure}, i.e. the union of outgoing neighborhoods for all infected nodes.  For most real world social networks, this implies the runtime scales slightly more than linearly with respect to the number of infections.

We can infer the model parameters for most contagions well inside a minute.  A large portion of real-world contagions in our dataset infects about 50-100 nodes, and rarely did the algorithm take more than 10 seconds to converge.  For larger contagions, some infecting thousands of nodes, the runtime was 5-10 minutes.

In all, we used the algorithm to fit the model to more than 18,000 real contagions and hundreds of synthetic contagions, and we never encountered convergence issues.

\section{Experiments}
\label{sec:experiments}
With our model well-defined and with an algorithm for inferring its parameters, we now apply it to real as well as synthetic data.  First, to establish the accuracy of the parameter inference algorithm, we fit our model to synthetic data. This allows for direct comparison of ground-truth to inferred parameters.  We examine a specific real-world case study to better illustrate the model.  Lastly, we run a series of large-scale experiments on the emergence of Twitter URLs. The model reveals the underlying dynamics of information emergence on Twitter.

\subsection{Experiments with synthetic data}
To test accuracy of the model parameter inference algorithm, we run a series of experiments on simulated data.

%
For each experiment, we first generate a large synthetic preferential attachment network.  We then choose values for $\eta(x)$, $\lambda_{ext}(t)$, and $\lambda_{int}(t)$.  At the start of the experiment, all nodes are uninfected.  Then, using a small discrete time step $\Delta t$ we march forward in time, and external exposures are sent to each node with probability $\lambda_{ext}(t) \cdot \Delta t$.  If a node becomes infected, it will transmit exactly one exposure to each of its outbound neighbors, and the time each outbound exposure takes to propagate is governed by $\lambda_{int}(t)\cdot \Delta t$.  With each exposure a node receives, we sample a binary random variable with bias $\eta(x)$ to determine whether the node will become infected upon that exposure.  Once the experiment is complete, the algorithm is given a set of node infection times, the underlying network, and  $\lambda_{int}(t)$, and its task is to infer $\eta(x)$ and $\lambda_{ext}(t)$.

\begin{figure}[t]
    \centering
    \subfigure[Event Profile $\lambda_{ext}(t)$]{\includegraphics[scale=.3]{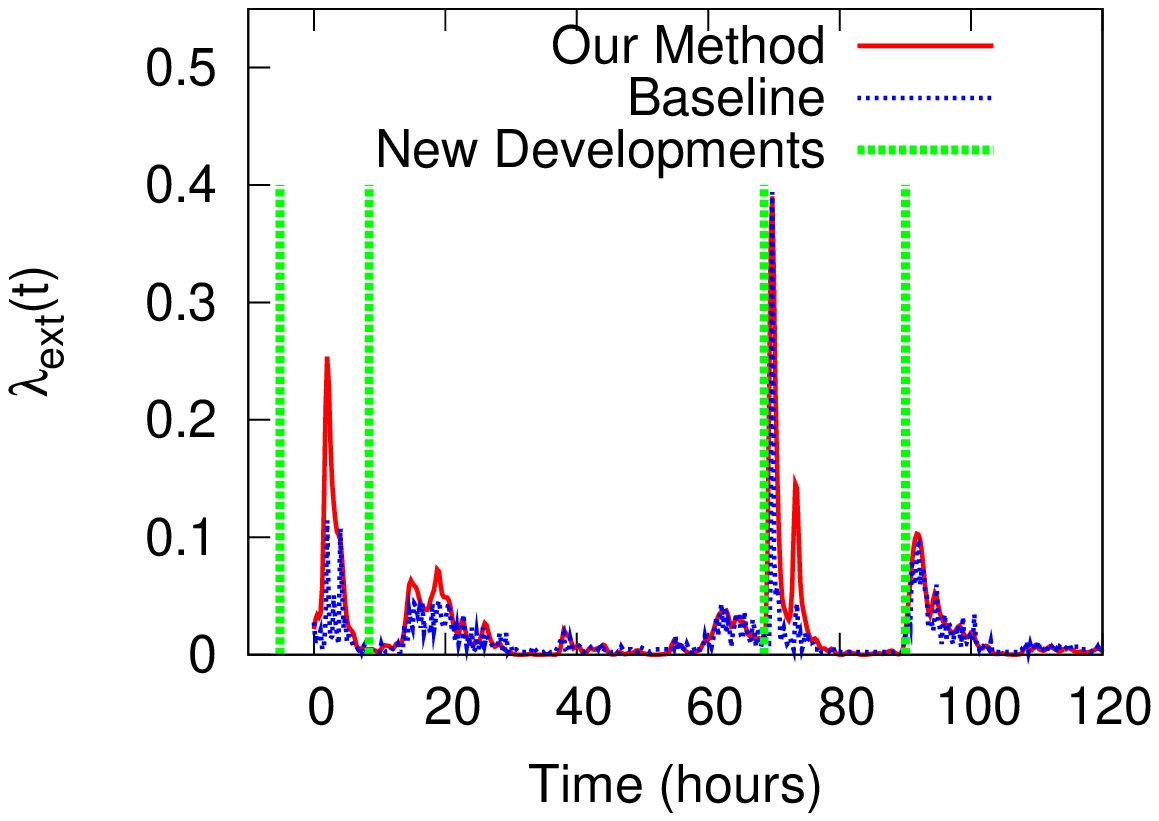}\label{fig:url_arizona_lamb}}
    \subfigure[Exposure Curve $\eta(x)$]{\includegraphics[scale=.3]{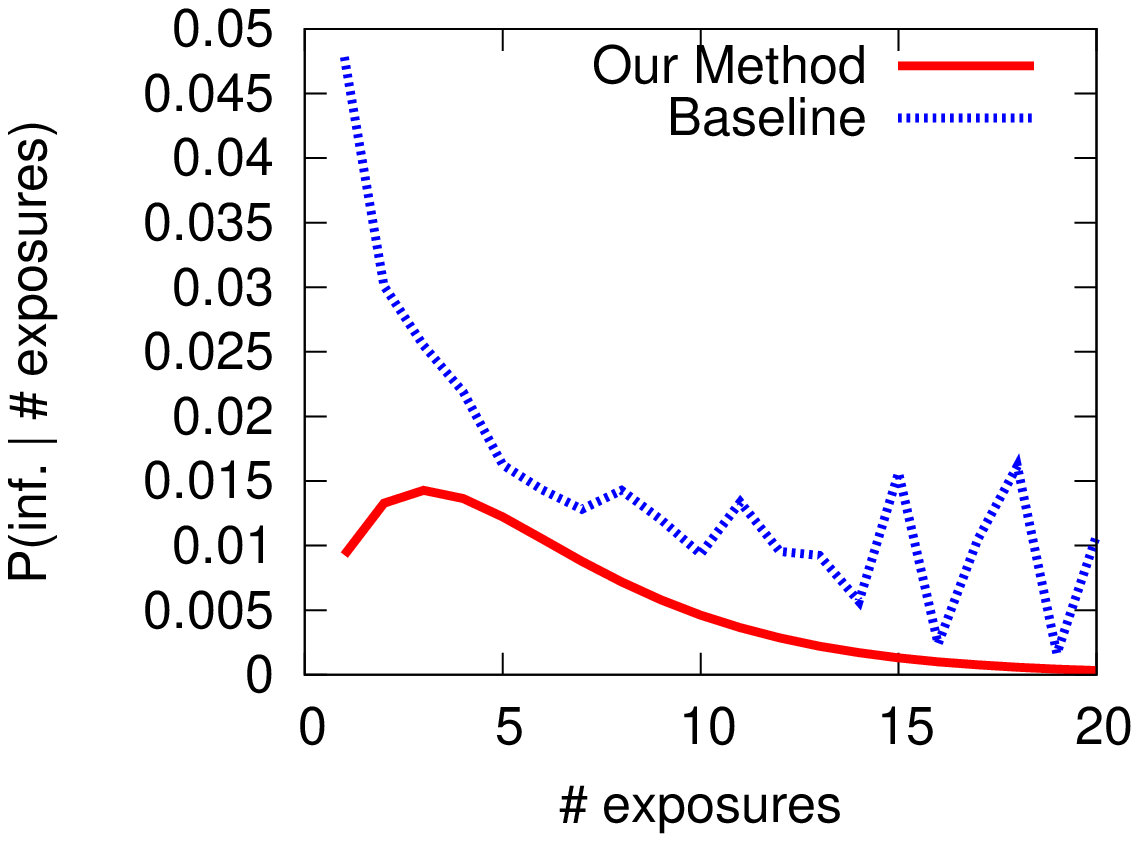}\label{fig:url_arizona_eta}}
\caption{ The model fitted to a single contagion representing URLs related to the Tucson, Arizona shootings.  The green vertical lines designate when four distinct developments related to the shooting event occurred.}
\label{fig:url_arizona}
    \vspace{-.2cm}
\end{figure}

\xhdr{Baselines}
We compared our algorithm against common sense baselines.  For inferring $\eta(x)$, we used the baseline of assuming internal exposures propagate immediately, and that all exposures originate internally.  Calculating $\eta(x_k)$ at each exposure count $x_k$ then boils down to counting the fraction of times a node becomes infected immediately after $x_k$ of its neighbors become infected. Note this is exactly the method of inferring $\eta(x)$ used in~\cite{romero11twitter}.  The baseline for inferring $\lambda_{ext}(t)$  uses the number of infections that occur for each unit of time in which none of the node's neighbors were previously infected.  We refer to these infections as external infections.  Since an externally infected node, by definition, has no infected neighbors, we know with certainty that all exposures the node received came from the event profile.  Therefore, the arrival of external infections over time should be indicative of the arrival of external exposures over time, \ie, the event profile.
This, however, only provides a {\em shape} (but not the scale) of the event profile, because without knowledge of the exposure curve $\eta(x)$, we do not know how many exposures it takes to typically cause an infection. Thus, the scale of the baseline $\lambda_{ext}(t)$ is usually 1 to 2 orders of magnitude larger.


\xhdr{Experimental results}
We ran many different combinations of network topologies, exposure curves, event profiles, and internal hazard functions.  Overall, we ran over 100 different combinations on networks of 75k+ nodes, and the algorithm not only performed consistently well but also did significantly better than the baselines.  We included the results of two such experiments in Fig. \ref{fig:syn_dat}.

For the first experiment, our algorithm is given a network and the data on Figure \ref{fig:syn_dat_1_all}. Based only on this information, it is able to infer data shown in Figures \ref{fig:syn_dat_1_ext} to  \ref{fig:syn_dat_1_eta}. These figures illustrate various aspects of the inferred profile of the external influence, \ie, the event profile, and exposure curve against the ground truth and the baselines.  For the event profile, not only is the scale of the baseline off by several orders of magnitude, but it also places the peak of the event profile far too early.  On the other hand, the event profile inferred using our algorithm very closely predicts the scale, shape, and the occurrence of the profile peak to the extent that the difference between the ground truth and inferred event profile is negligible.  The situation is the same for inferring the exposure curve in  Fig.~\ref{fig:syn_dat_1_eta}.  The inferred $\eta(x)$ almost exactly fits the ground truth, whereas the baseline overestimates the exposure curve by more than 50\%.


For the second experiment (Fig.~\ref{fig:syn_dat_1_ext}-\ref{fig:syn_dat_2_eta}), we used a very peculiar zig-zag ground-truth external influence profile (Fig.~\ref{fig:syn_dat_1_ext}), but the observations are still the same --- our model was able to infer all the quantities almost exactly.  The event profile inference shown in Fig.~\ref{fig:syn_dat_2_lamb} is very accurate.  It resolves each of the 10 peaks, while the baseline, besides being orders of magnitude off in scale, only detects 4 peaks.  We infer $\eta(x)$ almost exactly, as shown in Fig.~\ref{fig:syn_dat_2_eta}.


Note that even though we test the algorithm on synthetic data the fact that the model works well is not at all trivial. In particular, from the model fitting point of view the effects of internal and external influence are confounded and the model estimation procedure needs to separate them out. In particular, consider the contrast in the performance of the baseline approaches and the proposed model. Overall, these experiments demonstrate the robustness of the model and allow us to move to the experiments on real data.


\subsection{Experiments Using Real Data}

We now fit our model to a real data from the Twitter network.  We study the emergence of URLs on the Twitter network. URLs emerge by Twitter users mentioning them in their tweets (through tweeting or re-tweeting). Thus, URLs correspond to contagions, posting a tweet mentioning a particular URL corresponds to an infection event.


\xhdr{Twitter dataset}
To apply our model to a real-world information diffusion network, we collected {\em complete} Twitter data for January 2011, which consists of 3 billion  tweets. We focus on URLs that have been tweeted by at least 50 users as our contagions of study (we found that contagions smaller than 50 infections did not provide robust enough statistics).
For URLs that were shortened, we unshortened them and treated all URLs that point to the same web address as one contagion.  We restricted our focus to URLs in which we could classify as written in English.  To do this, we extracted natural text from the HTML of the URLs and then used a character sequenced-based classifier to determine their
language \cite{cavnar94langdetec}.  We also removed URLs that demonstrated blatant spamming behavior.  In all, this resulted in 18,186 different URLs.

We constructed the network over which these URLs propagate as follows.  First, we took the union of all users that tweeted at least one of these URLs.  Then, for each user in this set we used the Twitter API to extract a list of the users that they follow.  When one user follows another, he/she can see all of their tweets, include URLs that they post, and it is through this relationship that contagions spread on Twitter.  In all, this created a 1,087,033 node subgraph with 103,112,438 edges.  We focus our study on URLs as they clearly emerge due to external events.

For the internal hazard function $\lambda_{int}(t),$ empirical analysis indicates that $\lambda_{int}(t) \equiv \frac{0.14}{t }$, where $t$ is in hours, is a suitable choice.  This implies that the distribution of lag time between infections and exposures follows a power law with an exponent of 1.14.

\xhdr{A case study of the influence of external events}
We start our investigations on real data with an illustrative case study.  Using information diffusion, we aim to detect a sequence of external events that presumably caused bursts of activity on the Twitter network.

We examined the Tucson, Arizona shooting on January $8^{th}$ in which 6 people were killed and 14 others were injured, and among the injured was U.S. Congresswoman Gabrielle Giffords.  There were four key developments related to this event:
(1) the shooting occurs (Jan. 8, 10:10am),
(2) the Westboro Baptist Church announces plans to protest at the funerals of the victims (Jan. 9, 9:15am),
(3) Arizona Governor Jan Brewer signs emergency legislation blocking the protest. (Jan. 11, 9:24am), and
(4) an on-line ``Get Well Soon'' card is formed for Gabrielle Giffords that people can sign (Jan. 12, 6pm).

We collected all URLs that were tweeted at least 50 times that contained the word ``Giffords.'' We then gathered them into a single contagion.  Given that we aggregated four separate sub-stories we would expect that when we fit our model to the observed data, the event profile would coincide with developments related to the real-world event. Indeed this is the case as shown in Fig. \ref{fig:url_arizona}. 

The results of the model applied to the contagion are shown in Fig. \ref{fig:url_arizona}.  Additionally, the time of each of the 4 developments listed above is represented as a vertical green line in Fig \ref{fig:url_arizona_lamb}.  Our model clearly detects all four developments: each of them is followed by a spike in the event profile within 10 hours.  For the second two developments, the spikes in the event profile are immediate.  Also interesting is how the baseline event profile differs from the model's.  For example, immediately after the 3rd development (\ie, when the governor passed a new law) the model infers two spikes in $\lambda_{ext}(t)$ whereas the baseline records only one.  In response to the law being passed, many different groups began organizing counter protests to prevent the Westboro Baptist Church from interfering with the funerals.  This created a second influx of URLs from sources external to Twitter (Facebook groups, news sites, etc.), which was completely missed by the baseline.

\xhdr{Evaluation using Google Trends}
As a global alternative evaluation method we also performed the experiment where we extracted a set of mainstream media articles for which we were able to identify a single keyword $W$ that adequately describes them (\eg, swine flu for a BBC article on ``Increase in Northern Ireland swine flu cases''). For each $W$, we then queried Google Trends to obtain the number of worldwide search traffic of query $W$ over time. This served as a proxy for the activity of the external source.
We compared the L2 distance between the inferred event profile and the Google Trends ground-truth. Overall, we found that our model gives 30\% relative improvement in the L2 distance of the inferred event profile when compared to the naive event profile estimation.

\xhdr{External influence of different news categories} We now proceed to an aggregate analysis of event profiles and external influence of different category of news.
We identified 9 news sites that specify the article's category within the URL.  All together, we identified 1,929 URL's belonging to 11 different news categories.  We then fit our model each URL and infer the event profile as well as the exposure curve. For each news category, we then calculated the average $\rho_1$ which is the maximum probability of infection for the exposure curve, $\rho_2$ which is the number of exposures at which the URL is most infections, the duration or lifetime over which the event profile was inferred, and the number of expected total external exposures each node receives from the URL's event.

\addtocounter{table}{1}
\newcommand{\tabtopicresults}{\arabic{table}}
\begin{table*}[t]
\begin{tabular} { l r }
\imagetop{
\begin{minipage}{.61\textwidth}
\begin{minipage}{1\textwidth}
\begin{tabular} { l | c | c | c | c }
			&$\rho_1$			&$\rho_2$			&
\begin{minipage}{.07\textwidth}
\centering
{Duration (hours)}
\end{minipage}&
\begin{minipage}{.15\textwidth}
\centering
\% Ext. Exposures
\end{minipage}
\\
\hline
Politics (25)		& 0.0007 +/- 0.0001	& 4.59 +/- 0.76	& 51.24 +/- 16.66	& 47.38 +/- 6.12	\\
World (824)		& 0.0013 +/- 0.0000	& 2.97 +/- 0.10	& 43.54 +/- 2.94	& 26.07 +/- 1.19	\\
\begin{minipage}{.07\textwidth} Entertain. (117)\end{minipage}	& 0.0015 +/- 0.0002	& 3.52 +/- 0.28	& 89.89 +/- 16.13	& 17.87 +/- 2.51	\\
Sports (24)		& 0.0010 +/- 0.0003	& 4.76 +/- 0.83	& 87.85 +/- 38.03	& 43.88 +/- 6.97	\\
Health (81)		& 0.0016 +/- 0.0002	& 3.25 +/- 0.30	& 100.09 +/- 17.57	& 18.81 +/- 3.33	\\
Tech. (226)	& 0.0013 +/- 0.0001	& 3.00 +/- 0.16	& 83.05 +/- 8.73	& 18.36 +/- 1.80	\\
\begin{minipage}{.07\textwidth} Business (298)\end{minipage}		& 0.0015 +/- 0.0001	& 3.18 +/- 0.16	& 49.61 +/- 5.14	& 22.27 +/- 1.79	\\
\begin{minipage}{.07\textwidth}Science (106)\end{minipage}		& 0.0012 +/- 0.0002	& 4.06 +/- 0.30	& 135.28 +/- 16.19	& 20.53 +/- 2.78	\\
Travel (16)		& 0.0005 +/- 0.0001	& 2.33 +/- 0.29	& 151.73 +/- 39.70	& 39.99 +/- 6.60	\\
Art (32)		& 0.0006 +/- 0.0001	& 5.26 +/- 0.66	& 188.55 +/- 48.17	& 27.54 +/- 5.30	\\
Edu. (31)		& 0.0009 +/- 0.0001	& 3.77 +/- 0.51	& 130.53 +/- 38.63	& 21.45 +/- 6.40	\\
\end{tabular} \\ \, \end{minipage} \\
\begin{minipage}{1\textwidth}
\vspace{+.2cm}
\textbf{Table \arabic{table}: External Influence Model fit to news URL's belonging to various categories.  The values listed are the percent change that each topic's URLs are from the global average, for each parameter.}
\end{minipage} \end{minipage}}
&
\imagetop{
\begin{minipage}{.35\textwidth}
\begin{minipage}{1\textwidth}
\includegraphics[width=.95\textwidth]{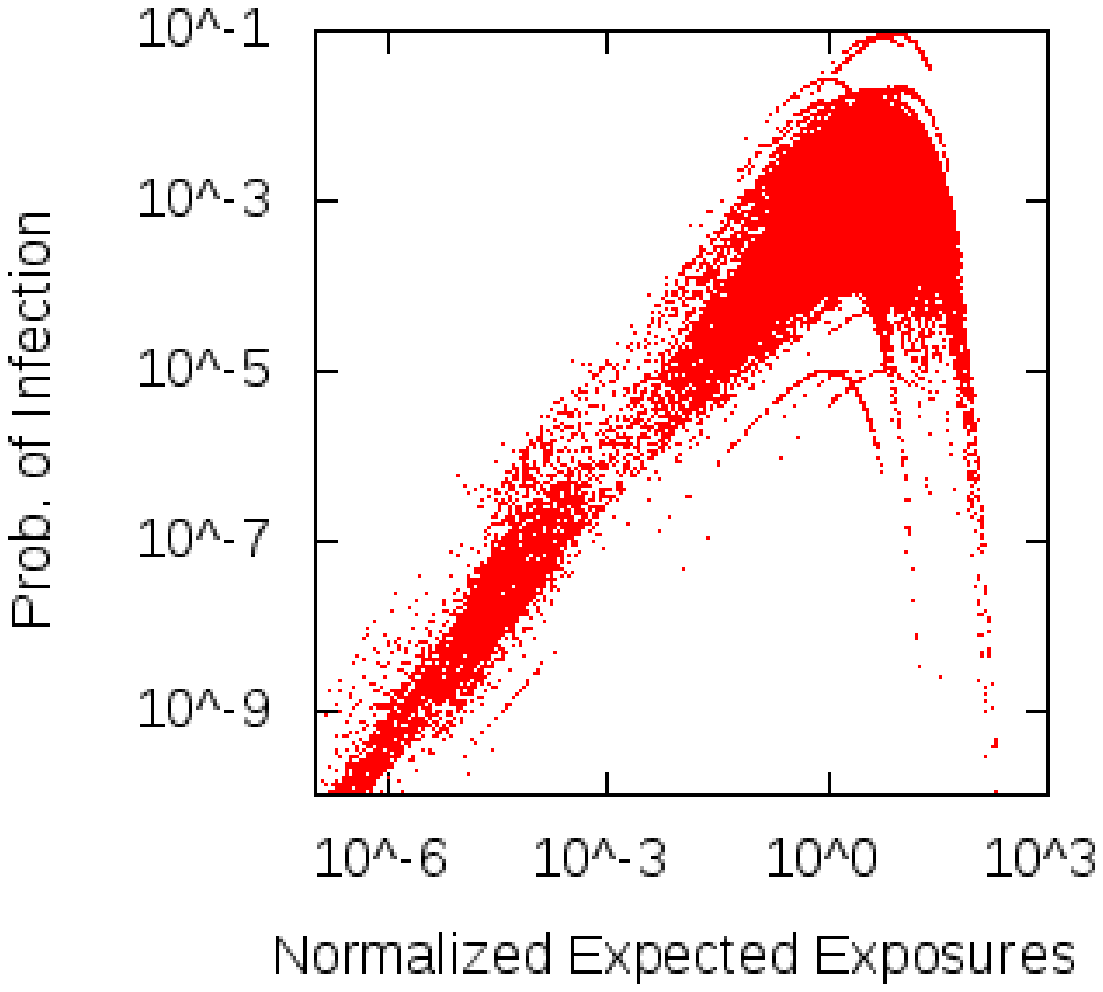}
\addtocounter{figure}{1}
\end{minipage}
\begin{minipage}{1\linewidth}
\textbf{Figure 5: An aggregation of the exposure curves for each URL.  Upon the infection of each user, the expected number of exposures received by the user divided by $\rho_2$ was plotted against inferred infection probability.}
\end{minipage}
\end{minipage}
}
\\
\end{tabular}

\end{table*}

The results are displayed in Table \tabtopicresults.  The average value of $\rho_1$ was 0.0013, $\rho_2$ was 3.21, the average duration of the contagions was 65.69 hours, and the average fraction of external infections was 23.94\%. In the first column, we show the maximum probability of infection for the exposure curve. Notice that Entertainment, Business, and Health appear to be the most infectious, where Art, Education, and Travel are the least infectious.  This seems reasonable as news articles about topics such as Art or Education would be less likely to be retweeted compared to Entertainment articles.  The second column describes upon which exposure the URL is most infectious.  World News, which is more time sensitive, reaches maximum infectiousness earlier compared to other topics.  After a user has received more than $\rho_2$ exposures, the probability of infection decreases, so it makes sense that these topics, which become irrelevant as time passes, reach this point sooner.  Contrast this with a topic like Art that is naturally less temporally sensitive.  Additionally, we learn that topics with a smaller $\rho_2$ tend to have shorter duration, and topics with a larger $\rho_2$ tend to have infections appear over a longer interval of time. Intuitively this makes sense as topics related to events (World, Business) get ``old'' sooner.

Lastly, the last column shows on average what percent of exposures came from external sources versus from within the network.  Politics appear to be the most externally driven topic, while Entertainment is the most internally driven.  This consistent with the fact that the 22 of the top 30 users followed on Twitter are entertainers.

\label{sec:findings}

\begin{figure*}[t]
\vspace{-.3cm}
    \centering
    \subfigure[Distribution of $\rho_1$]{\imagetop{\includegraphics[width=.33\linewidth]{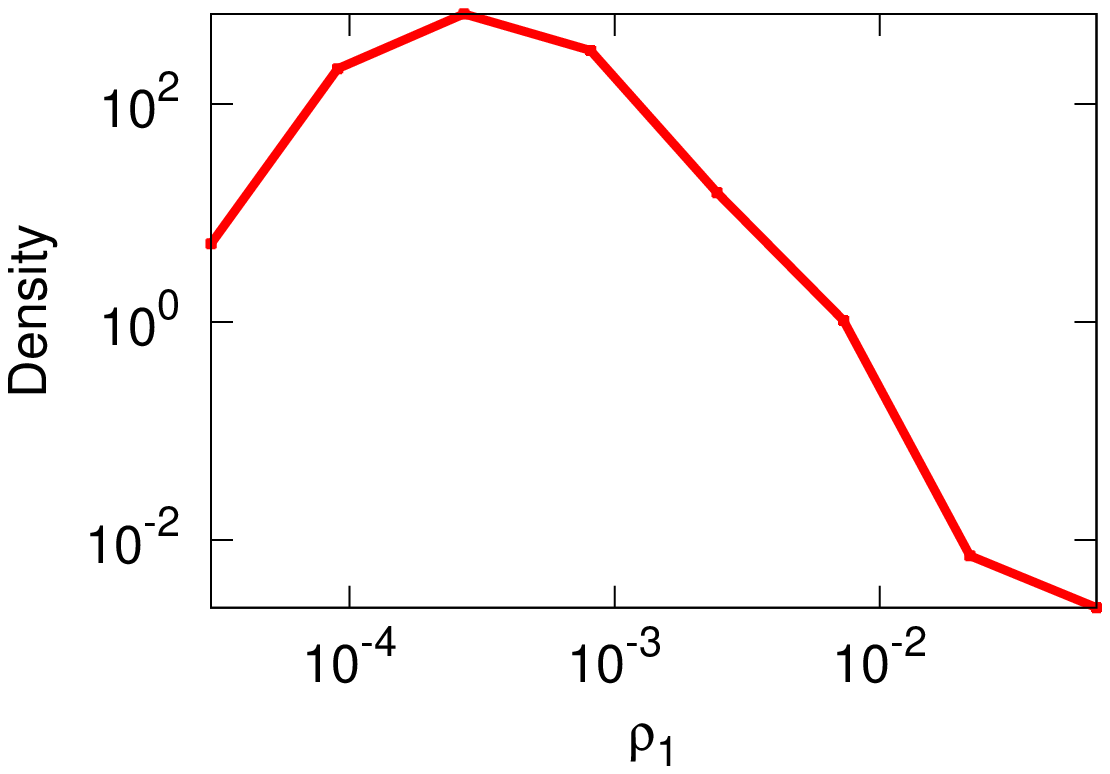}}\label{subfig:p1_dist}}
    \subfigure[Distribution of $\rho_2$]{\imagetop{\includegraphics[width=.33\linewidth]{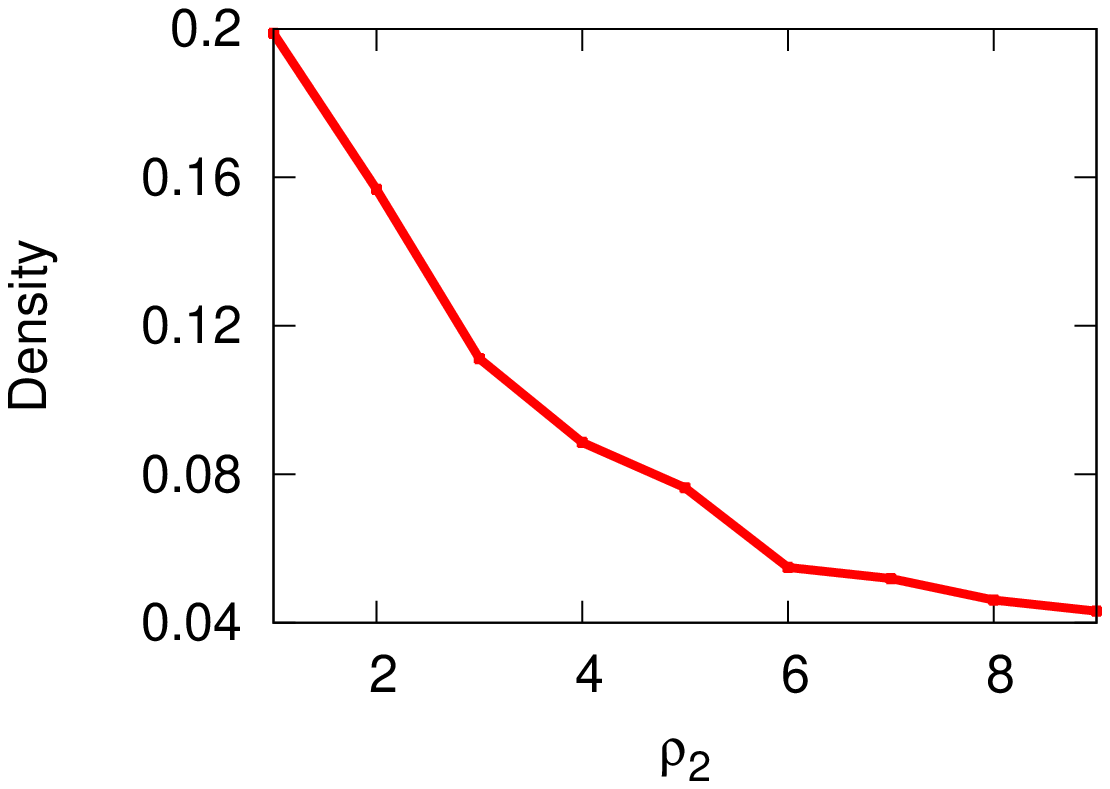}}\label{subfig:p2_dist}}
    \subfigure[Fraction of Internal Exposures]{\imagetop{\includegraphics[width=.33\linewidth]{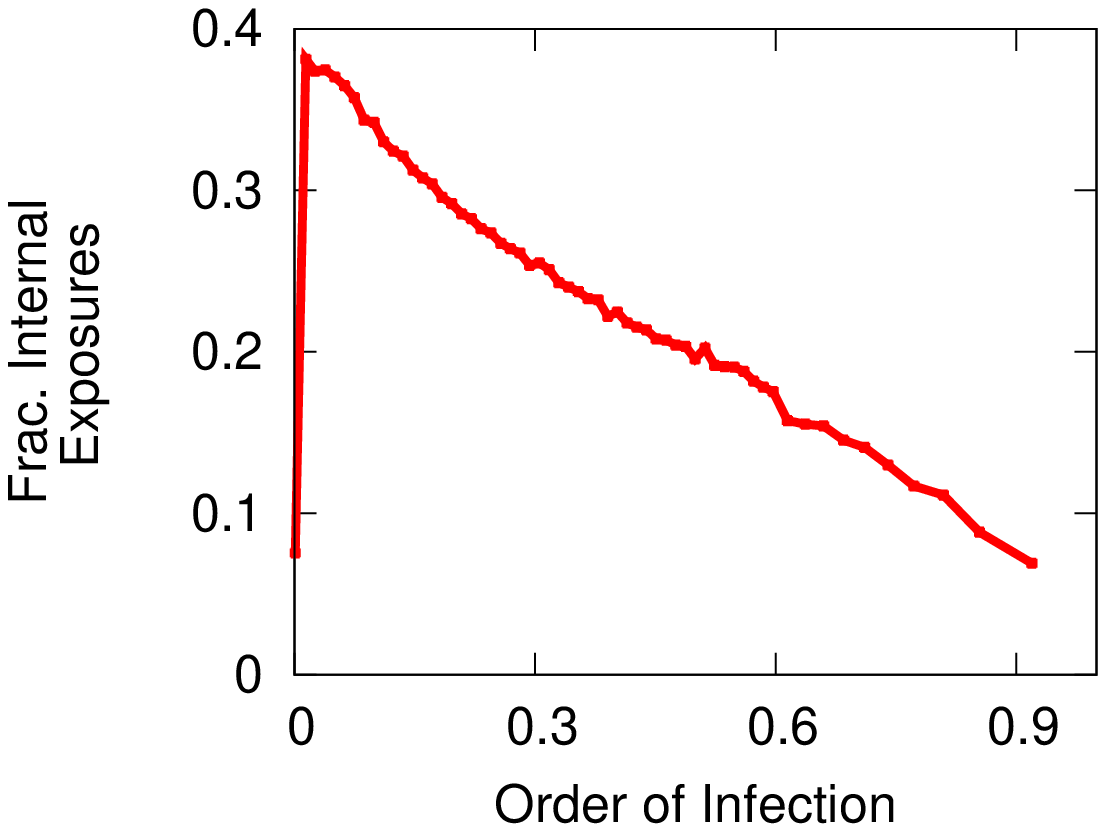}}\label{subfig:inforder}}
\vspace{-.5cm}
\caption{ (a)The distribution of the inferred $\rho_1$ (the max. of the exposure curve). (b)The distribution of the $\rho_2$ (the arg-max. of the exposure curve).  (c) For each infection of each URL, the fraction of users already infected was plotted against the inferred fraction of internal exposures the user received.  }
\label{fig:findings}
\vspace{-.3cm}
\end{figure*}
\xhdr{Global characteristics} The distributions for both the $\rho_1$ and $\rho_2$ exposure curve parameters inferred across the entire URL dataset can be found in Figures \ref{subfig:p1_dist}, \ref{subfig:p2_dist}.  Interesting is for how low the values of $\rho_1$ were inferred, with a mode on the order of .0005.  This implies that the people, at least Twitter users, are very selective about the ideas they adopt.  Additionally, most of the inferred $\rho_2$ parameters were small, with $\rho_2=1$ being the most common.  Recall that a smaller $\rho_2$ implies that the probability of infection begins to decrease with additional exposures sooner, and from this we see evidence that users quickly fatigue of most diffusing contagions.

Next, for each URL, we went through every user that was infected one by one.  For each user, we plotted the order of infection of the user in relation to all other infections versus the fraction of expected exposures the user received from internal sources, and the results can be found in Figure \ref{subfig:inforder}.  This plot demonstrates the interesting time dynamics at play.  On average, the first few users are infected almost purely externally, but then there is a surge in internal exposures.  As a result, the early infections are largely internally driven, but as the contagion continues to spread the infections are driven more and more by external influences.  This  initial surge in internally driven infections is also evident in the aggregated exposure curve, shown in Fig. 5.  Upon each infection, the expected number of exposures the user has received is recorded and divided by the inferred value of $\rho_2$.  This value shows how far along the node was in the exposure curve when the infection occurred, and the apex of the exposure curve occurs when it is equal to 1.  As one might expect, there is a high density of infections occurring at the apex.  What is interesting, however, is that there is also a dense group of infections happening early in the exposure curve at low probabilities.  This group is almost exclusively populated by internally infected users.

Finally, for each URL we calculated the expected number of exposures each user received during the emergence of the URL and what fraction of these exposures came from an external source.  Averaging across all URLs, we found that 71\% of all exposures came from internal sources within the network, while the other 29\% of the exposures were external.  We find this 29\% to be significant and clear evidence that external effects cannot be ignored.


\section{Conclusion}
\label{sec:conclusion}

Emergence of information has traditionally been solely modeled as a diffusion process in networks. However, we identified that only around 71\% of URL mentions on Twitter can be attributed to network effects, and the remaining 29\% of mentions seem to be due to the influence of external out-of-network sources. We then present a model in which information can reach a node via the links of the social network or through the influence of external sources. Applying the model to the emergence of URLs in the Twitter network demonstrated that our model can be used to infer the shape of influence functions as well as the effects of external sources on the information diffusion in networks. We should emphasize that our model does not only reliably capture the external influence but, as a consequence, also leads to a more accurate description of the real network diffusion process.

For future work it would be interesting to relax the assumption of uniform activity of the external source across all nodes of the network. Incorporating our model into methods for identifying ``influencers'' in networks~\cite{kempe03maximizing,bakshy11everyone,icwsm10cha} might be fruitful. Currently, phenomena we are observing are clearly taking place in aggregate. Ultimately, it will be interesting to pursue more fine-grained analyses as well, understanding how patterns of variation at the level of individuals contribute to the overall effects that we observe.

\xhdr{Acknowledgements}
This research has been supported in part by NSF
CNS-1010921,            
IIS-1016909,            
IIS-1149837,    
IIS-1159679,            
DARPA SMISC,
Albert Yu \& Mary Bechmann Foundation, Boeing, Allyes, Samsung,
Alfred P. Sloan and the Microsoft Faculty Fellowship.


\begin{thebibliography}{10}

\bibitem{adar05epidemics}
E.~Adar and L.~A. Adamic.
\newblock Tracking information epidemics in blogspace.
\newblock In {\em Web Intelligence}, pages 207--214, 2005.

\bibitem{aris08contagion}
A.~Anagnostopoulos, R.~Kumar, and M.~Mahdian.
\newblock Influence and correlation in social networks.
\newblock In {\em KDD '08}, 2008.

\bibitem{antoniades11urls}
D.~Antoniades, I.~Polakis, G.~Kontaxis, E.~Athanasopoulos, S.~Ioannidis, E.~P.
  Markatos, and T.~Karagiannis.
\newblock we.b: the web of short urls.
\newblock In {\em WWW '11}, 2011.

\bibitem{aral09contagion}
S.~Aral, L.~Muchnik, and A.~Sundararajan.
\newblock Distinguishing influence-based contagion from homophily-driven
  diffusion in dynamic networks.
\newblock {\em PNAS},
  106(51):21544--21549, 2009.

\bibitem{bakshy11everyone}
E.~Bakshy, J.~M. Hofman, W.~A. Mason, and D.~J. Watts.
\newblock Everyone's an influencer: quantifying influence on twitter.
\newblock In {\em WSDM'11}, 2011.

\bibitem{bennett-news-illusion}
L.~Bennett.
\newblock {\em News: {T}he Politics of Illusion}.
\newblock A. B. Longman (Classics in Political Science), seventh edition, 2006.

\bibitem{cavnar94langdetec}
W.~Cavnar and J.~Trenkle
\newblock{N-Gram-Based Text Categorization.}
\newblock{SDAIR, 94}

\bibitem{centola-complex-contagions}
D.~Centola and M.~Macy.
\newblock Complex contagions and the weakness of long ties.
\newblock {\em American Journal of Sociology}, 2007.

\bibitem{icwsm10cha}
M.~Cha, H.~Haddadi, F.~Benevenuto, and K.~P. Gummadi.
\newblock {Measuring User Influence in Twitter: The Million Follower Fallacy}.
\newblock In {\em ICWSM '10}, 2010.

\bibitem{cosley10influence}
D.~Cosley, D.~P. Huttenlocher, J.~M. Kleinberg, X.~Lan, and S.~Suri.
\newblock Sequential influence models in social networks.
\newblock In {\em ICWSM}, 2010.

\bibitem{crane08response}
R.~Crane and D.~Sornette.
\newblock Robust dynamic classes revealed by measuring the response function of
  a social system.
\newblock {\em PNAS},
  105(41):15649--15653, 2008.

\bibitem{johnson11hazard}
R.~Elandt-Johnson and N.~Johnson.
\newblock {\em Survival Models and Data Analysis}.
\newblock New York: John Wiley and Sons. 1980/1999.

\bibitem{rodriguez11}
M. G. Rodriguez, D. Balduzzi, and B. Sch\"{o}lkopf.
\newblock Uncovering the temporal dynamics of diffusion networks.
\newblock In {\em ICML '11}, 2011.

\bibitem{granovetter78threshold}
M.~S. Granovetter.
\newblock Threshold models of collective behavior.
\newblock {\em American Journal of Sociology}, 83(6):1420--1443, 1978.

\bibitem{hethcote00diseases}
H.~W. Hethcote.
\newblock The mathematics of infectious diseases.
\newblock {\em SIAM Review}, 42(4):599--653, 2000.

\bibitem{katz57twostep}
E.~Katz.
\newblock The {Two-Step} flow of communication: An {Up-To}-date report on an
  hypothesis.
\newblock {\em POQ},'57.

\bibitem{katz1955personal}
E.~Katz and P.~Lazarsfeld.
\newblock {\em {Personal influence: The part played by people in the flow of
  mass comm.}}.
\newblock Free Press, '55.

\bibitem{kempe03maximizing}
D.~Kempe, J.~M. Kleinberg, and E.~Tardos.
\newblock Maximizing the spread of influence through a social network.
\newblock In {\em KDD '03}.

\bibitem{kwak10twitter}
H.~Kwak, C.~Lee, H.~Park, and S.~Moon.
\newblock What is twitter, a social network or a news media?
\newblock In {\em WWW '10}, 2010.

\bibitem{lazarsfeld-peoples-choice}
P.~F. Lazarsfeld, B.~Berelson, and H.~Gaudet.
\newblock {\em The People's Choice: How the Voter Makes Up His Mind in a
  Presidential Campaign}.
\newblock Duell, Sloan, and Pearce, 1944.

\bibitem{jure09memes}
J.~Leskovec, L.~Backstrom, and J.~Kleinberg.
\newblock Meme-tracking and the dynamics of the news cycle.
\newblock In {\em KDD '09}, 2009.

\bibitem{jure07cascades}
J.~Leskovec, M.~McGlohon, C.~Faloutsos, N.~Glance, and M.~Hurst.
\newblock Cascading behavior in large blog graphs.
\newblock {\em SDM '07}.

\bibitem{rogers95diffusion}
E.~Rogers.
\newblock {\em Diffusion of Innovations}.
\newblock Free Press, 4th Ed,'95.

\bibitem{romero11twitter}
D.~M. Romero, B.~Meeder, and J.~M. Kleinberg.
\newblock Uncovering the Temporal Dynamics of Diffusion Networks
\newblock {\em WWW '11}.


\bibitem{shalizi10confounding}
C.~R. Shalizi and A.~C. Thomas.
\newblock Homophily and contagion are generically confounded in observational
  social network studies.
\newblock {\em Sociological Methods and Research}, 40, 2010.

\bibitem{steeg11epidem}
G.~Ver Steeg, R.~Ghosh, and K.~Lerman.
\newblock What Stops Social Epidemics?
\newblock In {\em ICWSM '11}.

\bibitem{strang98diffusion}
D.~Strang and S.~A. Soule.
\newblock Diffusion in organizations and social movements: From hybrid corn to
  poison pills.
\newblock {\em Annual Review of Sociology}, 24:265--290, 1998.

\bibitem{sun09gesundheit}
E.~Sun, I.~Rosenn, C.~Marlow, T.~Lento.
\newblock Gesundheit! modeling contagion through facebook.
\newblock {\em ICWSM '09}.

\bibitem{watts02cascades}
D.~J. Watts.
\newblock A simple model of global cascades on random networks.
\newblock {\em PNAS}

\bibitem{wu11twitter}
S.~Wu, J.~M. Hofman, W.~A. Mason, and D.~J. Watts.
\newblock Who says what to whom on twitter.
\newblock In {\em WWW '11}

\bibitem{dodds07influentials}
D.~J. Watts and P.~S. Dodds.
\newblock Influentials, networks, and public opinion formation.
\newblock {\em J. of Consumer Res.}, 34(4), 2007.

\end{thebibliography}



\clearpage

\end{document}